\documentclass[
 aps,
 prb,
 reprint,
 amsmath,amssymb,
 floatfix,
 nofootinbib,
]{revtex4-2}
\usepackage{graphicx} 
\usepackage{dcolumn}
\usepackage{bm}
\usepackage{braket}
\usepackage{multirow}
\usepackage[usenames]{color}
\usepackage{hyperref}
\usepackage{bbm}
\usepackage{mathtools}
\usepackage{capt-of}
\usepackage{orcidlink}

\hypersetup{
colorlinks = true, 
urlcolor = blue, 
linkcolor = blue, 
citecolor = blue
}

\newcommand{\bk}{\bm{k}}
\newcommand{\br}{\bm{R}}
\newcommand{\be}{\bm{e}}
\newcommand{\bq}{\bm{q}}
\newcommand{\unit}{\mathbbm{1}}
\newcommand{\Tr}{\mathrm{Tr}}

\newcommand{\mat}[1]{\bm{\mathrm{#1}}}

\begin{document}

\title{
Spin dynamics in disordered systems with generic spin--momentum couplings:\\ A matrix-rational formulation
}

\author{Masataka Kakoi}%\,\orcidlink{0009-0004-4243-3661}}
\email{kakoi@presto.phys.sci.osaka-u.ac.jp}
\affiliation{Department of Physics, The University of Osaka, Toyonaka, Osaka 560-0043, Japan}

\date{\today}

\begin{abstract}
    We develop a microscopic framework for the disorder-averaged density-matrix dynamics of pseudospin-$1/2$ systems with anisotropic and nonlinear spin--momentum couplings.
    Starting from the Diffuson, we introduce a matrix-valued rational approximation for its frequency dependence that does not rely on a low-frequency or weak-coupling expansion.
    The resulting pole equation is recast as a quadratic eigenvalue problem, from which the real-time evolution of the disorder-averaged density matrix is constructed.
    We apply the formulation to a two-dimensional electron gas with Rashba and linear and cubic Dresselhaus couplings, a $d$-wave altermagnet with Rashba coupling, and a three-dimensional double-Weyl model.
    Comparisons with direct numerical simulations demonstrate quantitative agreement for both the post-quench spin dynamics and spin relaxation times across a broad range of coupling strengths, extending beyond the diffusive regime.
    The matrix-rational formulation provides a unified description of ultrafast spin dynamics and long-time relaxation.
\end{abstract}

\maketitle

\section{Introduction}

Ultrafast spectroscopy has enabled real-time observations of nonequilibrium spin and carrier dynamics.
In optical pump--probe measurements, a delayed probe pulse tracks the spin polarization generated by optical excitation.
Time-resolved Kerr and Faraday rotation have been widely used to investigate spin relaxation and coherence in semiconductors~\cite{Wu_2010_review}.
Developments in time- and angle-resolved photoemission spectroscopy (TR-ARPES) have further enabled energy- and momentum-resolved measurements of transient electronic states~\cite{Boschini_2024_review}.
These techniques have revealed ultrafast spin dynamics and spin-dependent relaxation in materials with strong spin--orbit coupling (SOC)~\cite{Hsieh_2011,Cacho_2015,Sanchez-Barriga_2016,Clark_2022}.
Time-resolved experiments have also probed carrier dynamics in spin--momentum-locked topological surface states~\cite{Kuroda_2016,Reimann_2018,Reimann_2023}.
These experimental advances call for microscopic theories that resolve the coupled evolution of spin and momentum throughout the transient regime.

In the presence of disorder, spin evolution is governed by the interplay of precession and momentum scattering.
Because the SOC-induced effective field depends on momentum, even spin-independent scattering can alter spin precession and thereby relax spin polarization.
In the weak SOC regime, where the spin--orbit length is much longer than the scattering mean free path, the Dyakonov--Perel mechanism gives rise to motional narrowing, whereby stronger scattering slows spin relaxation~\cite{Dyakonov_1972}. 
This regime is captured by conventional spin diffusion theory.
When the SOC-induced splitting becomes comparable to or larger than the disorder energy scale, however, one must generally go beyond the diffusion approximation~\cite{Golub_2005,Glazov_2006,Glazov_2009,Grimaldi_2005,X-Liu_2011,X-Liu_2012,Hijano_2024,Sumita_arXiv}.

An exception to fast spin relaxation at strong SOC is the SU(2)-symmetric limit of balanced Rashba and linear Dresselhaus couplings, which supports a persistent spin helix (PSH)~\cite{Schliemann_2003,Bernevig_2006,Koralek_2009}.
Near this symmetry point, a long-lived in-plane spin component can coexist with fast oscillatory spin dynamics driven by the large spin splitting.
This separation of timescales highlights the importance of a unified description connecting the short-time transient response to long-time relaxation.
The possibility of protecting spin textures through crystal symmetry has also broadened interest in noncentrosymmetric materials~\cite{Tao_2018,Kilic_2025}.

Ultracold atoms also provide an ideal platform for exploring the real-time scattering dynamics of matter waves in random potentials.
Time-resolved measurements of momentum distributions directly track the scattering dynamics~\cite{Jendrzejewski_2012,Labeyrie_2012}.
The elastic scattering time can be tuned to the millisecond range, making the evolution on this timescale experimentally accessible~\cite{Richard_2019}.
Together with experimental realizations of synthetic SOC~\cite{Lin_2011,Huang_2016,Z-Wu_2016,ZY_Wang_2021,Leroux_2018,Hasan_2022}, these capabilities motivate theoretical studies of real-time dynamics in disordered spin--orbit-coupled systems~\cite{Mardonov_2015,Arabahmadi_2024,Oztas_2024,Kakoi_2024,Kakoi_2026a,Kakoi_2026b,Dornelas_2026}.

A common theoretical challenge emerging from ultrafast spectroscopy and cold-atom experiments is to describe spin--momentum dynamics across both short- and long-time regimes within a unified microscopic framework over a broad range of SOC strengths.
In Ref.~\cite{Kakoi_2026b}, such a description was obtained for spatially uniform SU(2) gauge fields, corresponding to spin--momentum couplings linear in momentum.
In this paper, we develop a systematic microscopic formulation of the disorder-averaged density-matrix dynamics for a broad class of spin--momentum couplings, including nonlinear momentum dependence.
This framework allows us to investigate how the strength and momentum dependence of the coupling govern post-quench evolution from short-time transients to long-time relaxation.

This paper is organized as follows.
In Sec.~\ref{sec:model}, we introduce the general pseudospin-$1/2$ model and the assumptions, thereby clarifying the scope of the proposed theory.
In Sec.~\ref{sec:microscopic_theory}, we formulate the disorder-averaged density-matrix dynamics in terms of the Diffuson.
We develop a matrix-valued rational approximation for its frequency dependence and recast the resulting pole equation as a finite-dimensional quadratic eigenvalue problem.
The resulting poles and associated residues are used to construct the density-matrix dynamics.
In Sec.~\ref{sec:demonstration}, we apply the resulting theory to representative models with nonlinear spin--momentum coupling and compare the predicted spin dynamics and relaxation times with numerical simulations.

\section{General model and assumptions}\label{sec:model}

We consider a $d$-dimensional disordered pseudospin-$1/2$ system described by
\begin{equation}
    H = H_0 + V,
\end{equation}
where $H_0$ is the clean Hamiltonian and $V$ is a scalar disordered potential.
Hereafter, we use ``spin'' to refer to the two-component internal (pseudospin) degree of freedom.
We assume a spatially $\delta$-correlated disorder,
\begin{equation}
    \overline{V(\bm{r})V(\bm{r}')}
    =
    \gamma_0\delta(\bm{r}-\bm{r}'),
    \label{eq:disorder_correlation}
\end{equation}
where $\overline{(\cdots)}$ denotes disorder average and $\gamma_0$ characterizes the disorder strength.
The clean Hamiltonian is taken in the generic form
\begin{equation}
    H_0(\bk)
    =
    \varepsilon_0(\bk)\unit_2
    +
    \br(\bk)\cdot\bm{\sigma},
    \label{eq:Hamiltonian}
\end{equation}
where $\bm{\sigma}=(\sigma_x,\sigma_y,\sigma_z)$ denotes the Pauli matrices and
$\br(\bk)$ is a real vector-valued function of momentum.
The two energy branches of $H_0(\bk)$ are
\begin{equation}
    \mathcal{E}_s(\bk)
    =
    \varepsilon_0(\bk)
    +
    s|\br(\bk)|,
    \qquad
    s=\pm.
    \label{eq:branch_energy}
\end{equation}
For convenience, we define
\begin{equation}
    \hat{\br}_{\bk}
    =
    \frac{\br(\bk)}{|\br(\bk)|}.
\end{equation}
The projection operator onto branch $s$ is then
\begin{equation}
    P_s(\bk)
    =
    \frac{1}{2}
    \left[
        \unit_2
        +
        s\hat{\br}_{\bk}\cdot\bm{\sigma}
    \right].
    \label{eq:projector}
\end{equation}
Throughout this work, we assume that the Hamiltonian satisfies the following condition.
Specifically, there exists a set $\mathcal{G}$ of transformations $\bk\mapsto\mathfrak{g}\bk$ such that
\begin{equation}
    \varepsilon_0(\mathfrak{g}\bk)=\varepsilon_0(\bk),
    \quad
    |\br(\mathfrak{g}\bk)|=|\br(\bk)|,
    \quad
    \sum_{\mathfrak{g}\in\mathcal{G}}\br(\mathfrak{g}\bk)=0.
    \label{eq:symmetry_condition}
\end{equation}
In other words, the symmetry-related momenta have the same branch energies, while their spin dependent fields sum to zero.
The usual spin-$1/2$ time-reversal-symmetric case,
\begin{equation}
    \varepsilon_0(-\bk)=\varepsilon_0(\bk),
    \qquad
    \br(-\bk)=-\br(\bk),
    \label{eq:TRS_condition}
\end{equation}
is the simplest realization of Eq.~\eqref{eq:symmetry_condition}.

The disorder-averaged retarded Green's function for branch $s$ is written as
\begin{equation}
    g_s(\bk,E)
    =
    \frac{1}{
        E-\mathcal{E}_s(\bk)+i\frac{\hbar}{2\tau}
    },
    \label{eq:branch_green}
\end{equation}
where $\tau$ is the elastic scattering time and the real part of the self-energy is absorbed into $\mathcal{E}_s$.
For the $\delta$-correlated disorder considered here, the Born approximation gives
\begin{equation}
    \frac{\hbar}{\tau(E)}
    =
    \pi\nu(E)\gamma_0,
    \label{eq:tau_gamma}
\end{equation}
where $\nu(E)$ denotes the density of states including both branches~\cite{Kakoi_2026a}.
Under the symmetry condition in Eq.~\eqref{eq:symmetry_condition}, the spin-dependent part of the Born self-energy vanishes upon momentum integration, so that the self-energy is proportional to the identity in spin space.
Using the projection operators in Eq.~\eqref{eq:projector}, the disorder-averaged retarded Green's function in the pseudospin basis can be written as
\begin{equation}
    \overline{G}(\bk,E)
    =
    \sum_{s=\pm}
    g_s(\bk,E)P_s(\bk).
    \label{eq:G_projector}
\end{equation}
Expanding the projection operators, we obtain
\begin{align}
    \overline{G}(\bk,E)
    &=
    \frac{g_+(\bk,E)+g_-(\bk,E)}{2}\unit_2
    \notag\\
    &\quad+
    \frac{g_+(\bk,E)-g_-(\bk,E)}{2}
    \hat{\br}_{\bk}\cdot\bm{\sigma}.
    \label{eq:G_expanded}
\end{align}

\section{Microscopic theory of disorder-averaged dynamics}
\label{sec:microscopic_theory}

Spin dynamics in spin--orbit-coupled disordered systems can be formulated in terms of the Diffuson.
Its poles characterize the decay and oscillation of disorder-averaged densities, with those in the spin sector determining the spin relaxation modes and their characteristic timescales~\cite{Burkov_2004a,Burkov_2004b,Kakoi_2026b}.
In Sec.~\ref{sec:diffuson_formalism}, we formulate the disorder-averaged density-matrix for a generic pseudospin-$1/2$ Hamiltonian in terms of the Diffuson.
In Sec.~\ref{sec:rational_approx}, we develop a matrix-valued rational approximation for the frequency dependence of the Diffuson, without restricting the SOC to a specific momentum dependence or to the weak SOC diffusive regime.
In Sec.~\ref{sec:QEP}, we recast the resulting pole equation as a quadratic eigenvalue problem and then linearize it into a generalized eigenvalue problem.
In Sec.~\ref{sec:pole_residues}, we derive the residue matrices and show how the pole expansion enters the disorder-averaged density-matrix dynamics.
Finally, Sec.~\ref{sec:computational_procedure} summarizes the computational procedure.

\subsection{Diffuson formalism}
\label{sec:diffuson_formalism}

To connect the Diffuson with the disorder-averaged spin dynamics, we first recall its relation to the density matrix.
The disorder-averaged density matrix $\hat{\varrho}(t)$ is obtained from the intensity propagator as~\cite{Kuhn_2007,Kakoi_2026b}
\begin{align}
    \varrho_{\alpha\beta}(\bk,\bq,t)
    &=
    \sum_{\gamma,\delta}
    \int\!\frac{\hbar\,d\omega}{2\pi}
    \int\!\frac{dE}{2\pi}
    \int_{\bk'}
    e^{-i\omega t}
    \notag\\
    &\quad\times
    \Phi_{\alpha\beta,\gamma\delta}
    (\bk,\bk'\!,\bq,E,\omega)
    \varrho_{\gamma\delta}(\bk'\!,\bq,0).
    \label{eq:density_matrix_from_Phi}
\end{align}
Here, 
\begin{align}\label{eq:def_Phi}
    &\Phi_{\alpha\beta,\gamma\delta}(\bk,\bk'\!,\bq,E,\omega)\notag\\
    &= \overline{
    \braket{\bk_+,\alpha|
    \hat{G}\!\left(E_+\right)
    |\bk_+',\gamma}
    \braket{\bk_-',\delta|
    \hat{G}^\dag\!\left(E_-\right)
    |\bk_-,\beta}}
\end{align}
is the intensity propagator.
For convenience, we define
\begin{equation}
    \bk_\pm
    :=
    \bk\pm\frac{\bq}{2},
    \quad
    E_\pm
    :=
    E\pm\frac{\hbar\omega}{2},
    \quad
    \int_{\bk}
    :=
    \int\!\frac{d^d\bk}{(2\pi)^d}.
    \label{eq:kEpm}
\end{equation}
Within the standard diagrammatic theory~\cite{Akkermans_2007_book}, the intensity propagator is approximated by retaining the Drude--Boltzmann ($\Phi^0$), Diffuson ($\Phi^\mathrm{D}$), and Cooperon ($\Phi^\mathrm{C}$) contributions:
\begin{equation}
    \Phi \simeq \Phi^0 + \Phi^\mathrm{D} + \Phi^\mathrm{C}.
\end{equation}
The Drude--Boltzmann contribution is given by the product of two disorder-averaged Green's functions.\footnote{
The Drude--Boltzmann contribution in the $\pm$ basis is
\[
    \Phi^0_{ss'}(\bk,\bk'\!,\bq,E,\omega) = (2\pi)^d\delta(\bk-\bk')\,
    g_s(\bk_+,E_+)g_{s'}^*(\bk_-,E_-).
\]
}
The Diffuson accounts for multiple scattering through the ladder diagrams.
Since $\Phi^0$ can be evaluated straightforwardly~\cite{Kakoi_2024}, we focus on the Diffuson contribution.
The Cooperon describes interference between time-reversed scattering paths and gives rise to weak (anti)localization, which is not considered in this work.
Nevertheless, much of the following analysis can also be applied to the Cooperon by replacing the particle--hole propagator with the corresponding particle--particle propagator.

For scalar, spatially $\delta$-correlated disorder, the Diffuson contribution to the intensity propagator is
\begin{equation}
\begin{aligned}
    &\Phi^\mathrm{D}(\bk,\bk'\!,\bq,E,\omega)
    \\
    &=
    \left[
        \overline{G}(\bk_+,E_+)
        \otimes
        \overline{G}{}^{*}(\bk_-,E_-)
    \right]
    \\
    &\quad\times
    \Gamma^\mathrm{D}(\bq,E,\omega)
    \left[
        \overline{G}(\bk'_+,E_+)
        \otimes
        \overline{G}{}^{*}(\bk'_-,E_-)
    \right],
\end{aligned}
\label{eq:PhiD_residue}
\end{equation}
where the Diffuson vertex is given by the ladder sum~\cite{Akkermans_2007_book},
\begin{equation}
    \Gamma^\mathrm{D}(\bq,E,\omega)
    =
    \gamma_0
    \left[
        \unit_4
        -
        \gamma_0\Pi^\mathrm{D}(\bq,E,\omega)
    \right]^{-1},
    \label{eq:Diffuson}
\end{equation}
with the particle--hole propagator
\begin{equation}
    \Pi^\mathrm{D}_{\alpha\beta,\gamma\delta}
    (\bq,E,\omega)
    =
    \int_{\bk}
    \overline{G}_{\alpha\gamma}(\bk_+,E_+)
    \overline{G}{}^{*}_{\beta\delta}(\bk_-,E_-).
    \label{eq:PiD_spin_basis}
\end{equation}
The poles of $\Gamma^\mathrm{D}$, which govern the disorder-averaged dynamics, are determined by
\begin{equation}
    \det\!\left(
        \unit_4
        -
        \gamma_0\Pi^\mathrm{D}
    \right)
    =
    0.
    \label{eq:Diffuson_pole_general}
\end{equation}
Because the particle--hole propagator $\Pi^\mathrm{D}$ depends nonlinearly on the complex frequency $\omega$, determining the pole frequencies amounts to solving a nonlinear eigenvalue problem~\cite{Guttel_2017_review}.

It is convenient to represent the Diffuson in the orthonormal Pauli basis\footnote{
A $2\times2$ density matrix can be expanded as
\[
    \varrho = \varrho_0 T_0 + \varrho_x T_x + \varrho_y T_y + \varrho_z T_z,
\]
and can therefore be represented by the four-component vector
$(\varrho_0,\varrho_x,\varrho_y,\varrho_z)$.
}
\begin{equation}
    T_0=\frac{\unit_2}{\sqrt{2}},
    \quad
    T_x=\frac{\sigma_x}{\sqrt{2}},
    \quad
    T_y=\frac{\sigma_y}{\sqrt{2}},
    \quad
    T_z=\frac{\sigma_z}{\sqrt{2}}.
    \label{eq:Pauli_basis}
\end{equation}
Here, $T_0$ represents the density component, while $T_{x,y,z}$ represent the three spin components.
In this basis, the propagator is
\begin{equation}
    \Pi^\mathrm{D}_{ij}
    =
    \sum_{\alpha,\beta,\gamma,\delta}
    (T_i)^*_{\alpha\beta}
    \Pi^\mathrm{D}_{\alpha\beta,\gamma\delta}
    (T_j)_{\gamma\delta},
    \label{eq:chsp_transform}
\end{equation}
or equivalently,
\begin{equation}
    \Pi^\mathrm{D}_{ij}(\bq,E,\omega)
    =
    \int_{\bk}
    \Tr\!\left[
        T_i^\dagger
        \overline{G}(\bk_+,E_+)
        T_j
        \overline{G}{}^\dagger(\bk_-,E_-)
    \right].
    \label{eq:PiD_chsp_trace}
\end{equation}
The poles associated with the spin components of the Diffuson determine the spin relaxation modes and their characteristic times.
We now focus on the momentum-diagonal density matrix, corresponding to $\bq=\bm{0}$.
Under the symmetry condition in Eq.~\eqref{eq:symmetry_condition}, $g_s(\mathfrak{g}\bk,E)=g_s(\bk,E)$ for all $\mathfrak{g}\in\mathcal{G}$ [see Eq.~\eqref{eq:branch_green}].
Consequently, terms linear in $\hat{\br}_{\bk}$ cancel upon momentum integration (see Appendix~\ref{app:diffuson_derivation}).
Therefore,
\begin{equation}
    \Pi^\mathrm{D}_{0i}
    =
    \Pi^\mathrm{D}_{i0}
    =
    0,
    \qquad
    i=x,y,z,
    \label{eq:density_spin_decoupling}
\end{equation}
and the propagator takes the block-diagonal form
\begin{equation}
    \Pi^\mathrm{D}
    =
    \Pi^\mathrm{D}_{00}
    \oplus
    \Pi^\mathrm{D}_{\mathrm{sp}}.
    \label{eq:density_spin_blocks}
\end{equation}
Hereafter, the argument $\bq=\bm{0}$ is left implicit.

Substituting Eq.~\eqref{eq:G_projector} into Eq.~\eqref{eq:PiD_chsp_trace} and using the Pauli-matrix algebra, we obtain the following density and spin components.
The details of the derivation are given in Appendix~\ref{app:diffuson_derivation}.
The density component is
\begin{equation}
    \Pi^\mathrm{D}_{00}(E,\omega) =
    \frac{1}{2}
    \int_{\bk}
    \mathcal{P}_{\parallel}(\bk,E,\omega),
    \label{eq:PiD_density}
\end{equation}
while the spin block can be decomposed into two contributions,
\begin{align}
    \Pi^\mathrm{D}_{\mathrm{sp}}
    &=\Pi^\mathrm{D}_{\mathrm{sp},\parallel}
    + \Pi^\mathrm{D}_{\mathrm{sp},\perp},
    \label{eq:PiD_spin}
\end{align}
where
\begin{align}
    \bigr[\Pi^\mathrm{D}_{\mathrm{sp},\parallel}\bigr]_{ij}
    &=
    \frac{1}{2}
    \int_{\bk} \mathcal{P}_{\parallel}(\bk,E,\omega)
    \hat{R}_{\bk,i}\hat{R}_{\bk,j}, 
    \label{eq:PiD_spin_parallel} \\
    \bigr[\Pi^\mathrm{D}_{\mathrm{sp},\perp}\bigr]_{ij}
    &=
    \frac{1}{2}
    \int_{\bk}
    \mathcal{P}_{\perp}(\bk,E,\omega)
    \left(
    \delta_{i,j} -
    \hat{R}_{\bk,i}\hat{R}_{\bk,j}
    \right).
    \label{eq:PiD_spin_perp}
\end{align}
Here,
\begin{align}
    \mathcal{P}_{\parallel}(\bk,E,\omega)
    &=
    \sum_{s=\pm}
    g_s(\bk,E_+)
    g_s^*(\bk,E_-),
    \label{eq:Pparallel}
    \\
    \mathcal{P}_{\perp}(\bk,E,\omega)
    &=
    \sum_{s=\pm}
    g_s(\bk,E_+)
    g_{-s}^*(\bk,E_-).
    \label{eq:Pperp}
\end{align}
The subscripts $\parallel$ and $\perp$ denote the same-branch and different-branch products, respectively.

\subsection{Matrix-valued rational approximation}
\label{sec:rational_approx}

To determine the spin relaxation modes from the poles of the Diffuson, we seek an approximate rational representation of its frequency dependence that remains applicable beyond the diffusive regime.
We first simplify $\mathcal{P}_{\parallel}$ and $\mathcal{P}_{\perp}$ in Eqs.~\eqref{eq:Pparallel} and \eqref{eq:Pperp}.
Following Ref.~\cite{Kakoi_2026b}, we evaluate the spectral weight at the central energy $E$, while retaining the full dependence on $\omega\tau$ and the branch splitting.
This gives
\begin{align}
    \mathcal{P}_{\parallel}(\bk,E,\omega)
    &\simeq
    \frac{2\mathcal{A}(\bk,E)}{\gamma_0\nu(E)}
    \frac{1}{z_\omega},
    \label{eq:Pparallel_approx}
    \\
    \mathcal{P}_{\perp}(\bk,E,\omega)
    &\simeq
    \frac{2\mathcal{A}(\bk,E)}{\gamma_0\nu(E)}
    \frac{z_\omega}{z_\omega^2+\Delta_{\bk}^2},
    \label{eq:Pperp_approx}
\end{align}
where
\begin{equation}
    z_\omega=1-i\omega\tau,
\end{equation}
and the dimensionless branch splitting is defined as
\begin{equation}
    \Delta_{\bk}
    =
    \frac{\tau}{\hbar}
    \left[
        \mathcal{E}_+(\bk)-\mathcal{E}_-(\bk)
    \right]
    =
    \frac{2\tau|\br(\bk)|}{\hbar}.
    \label{eq:z_Delta}
\end{equation}
The spectral function is defined by
\begin{equation}
    \mathcal{A}(\bk,E)
    =
    -\frac{1}{\pi}
    \operatorname{Im}\!
    \left[
        \Tr_{\rm int}\overline{G}(\bk,E)
    \right],
    \label{eq:spectral_DOS}
\end{equation}
where $\Tr_{\rm int}$ denotes the trace over the internal spin degree of freedom.

With this approximation, the density component becomes
\begin{equation}
    \gamma_0\Pi^\mathrm{D}_{00}(E,\omega)
    =
    \frac{1}{z_\omega}.
    \label{eq:PiD_density_approx}
\end{equation}
$\Pi^\mathrm{D}_{{\rm sp},\parallel}$ in Eq.~\eqref{eq:PiD_spin_parallel} can likewise be written as
\begin{equation}
    \gamma_0
    \Pi^\mathrm{D}_{{\rm sp},\parallel}(E,\omega)
    =
    \frac{1}{z_\omega}\mat{C},
\end{equation}
where $\mat{C}$ is a $3\times3$ real matrix defined by
\begin{equation}
    \mat{C}_{ij}
    =
    \frac{1}{\nu(E)}
    \int_{\bk}
    \mathcal{A}(\bk,E)
    \hat{R}_{\bk,i}\hat{R}_{\bk,j}.
    \label{eq:C_definition}
\end{equation}
For weak disorder, the momentum integral can be evaluated within the on-shell approximation~\cite{Kakoi_2026b} by replacing it with an average over the constant-energy shell at $E=E_0$,
\begin{equation}
    \mat{C}
    =
    \left\langle
        \hat{\br}_{\bk}\hat{\br}_{\bk}^{\top}
    \right\rangle_{E_0},
    \label{eq:C_FS}
\end{equation}
where $\top$ denotes matrix transposition.
By construction, $\mat{C}$ is positive semidefinite with $\Tr\,\mat{C}=1$.

The different-branch contribution $\Pi^\mathrm{D}_{{\rm sp},\perp}$ in Eq.~\eqref{eq:PiD_spin_perp} is more involved because it contains
\begin{equation}
    \frac{z_\omega}
    {z_\omega^2+\Delta_{\bk}^2}
    \left(
        \unit_3
        -
        \hat{\br}_{\bk}\hat{\br}_{\bk}^{\top}
    \right).
\end{equation}
In general, the $\bk$ dependence of $\Delta_{\bk}$ is correlated with that of $\hat{\br}_{\bk}$, and the corresponding momentum integral therefore cannot be reduced to a single scalar spin--orbit scale.
The approximation developed below follows the strategy in Ref.~\cite{Kakoi_2026b}, where rational-function ansatze were used to capture the frequency dependence of the Diffuson and Cooperon beyond the diffusion approximation treating the SOC nonperturbatively.
For the uniform SU(2) gauge field considered there, the spin sector decouples into three independent scalar channels, each governed by a cubic pole equation.
For the more general Hamiltonian considered here, such a decoupling is not generally possible. 
We therefore retain the full $3\times3$ matrix structure of the particle--hole propagator and seek a matrix-valued generalization.
We assume $\Pi^\mathrm{D}_{{\rm sp},\perp}$ in Eq.~\eqref{eq:PiD_spin_perp} to take the form
\begin{equation}
    \gamma_0
    \Pi^\mathrm{D}_{{\rm sp},\perp}(E,\omega)
    \simeq
    z_\omega
    \mat{M}_0^{1/2}
    \left(
        z_\omega^2\unit_3+\mat{L}
    \right)^{-1}
    \mat{M}_0^{1/2}.
    \label{eq:matrix_rational_approx}
\end{equation}
Here,
\begin{equation}
\begin{aligned}
    \mat{M}_0
    &=
    \left\langle
        \unit_3-\hat{\br}_{\bk}\hat{\br}_{\bk}^{\top}
    \right\rangle_{\!E_0}
    =
    \unit_3-\mat{C},
    \\
    \mat{M}_1
    &=
    \left\langle
        \frac{
            \unit_3-\hat{\br}_{\bk}\hat{\br}_{\bk}^{\top}
        }{
            1+\Delta_{\bk}^2
        }
    \right\rangle_{\!E_0},
    \label{eq:M0M1}
\end{aligned}
\end{equation}
and\footnote{
To include cases with $\operatorname{rank}\mat{M}_0<3$,
Eq.~\eqref{eq:L_definition} is understood on the support of $\mat{M}_0$, with $\mat{M}_1^{-1}$ denoting the inverse restricted to this subspace and $\unit_3$ the corresponding identity.
}
\begin{equation}
    \mat{L}
    =
    \mat{M}_0^{1/2}
    \mat{M}_1^{-1}
    \mat{M}_0^{1/2}
    -
    \unit_3.
    \label{eq:L_definition}
\end{equation}
By construction, $\mat{M}_0$, $\mat{M}_1$, and $\mat{L}$ are positive semidefinite.\footnote{\begin{minipage}[t]{\linewidth}
Since
\[
    0
    \preceq
    \frac{\unit_3-\hat{\br}_{\bk}\hat{\br}_{\bk}^{\top}}
         {1+\Delta_{\bk}^2}
    \preceq
    \unit_3-\hat{\br}_{\bk}\hat{\br}_{\bk}^{\top},
\]
we have $0\preceq\mat{M}_1\preceq\mat{M}_0$.
Moreover, since $1/(1+\Delta_{\bk}^2)>0$, $\mat{M}_0$ and $\mat{M}_1$ have the same kernel and are therefore both positive definite on the support of $\mat{M}_0$.
Defining $\mat{X} = \mat{M}_0^{-1/2} \mat{M}_1 \mat{M}_0^{-1/2}$, we obtain $0\prec\mat{X}\preceq\unit_3$, and hence $\mat{L}=\mat{X}^{-1}-\unit_3\succeq0$.
\end{minipage}}
As discussed later, this positive semidefiniteness guarantees the absence of unphysical spin modes within the rational approximation.
The form of Eq.~\eqref{eq:matrix_rational_approx} is chosen to reproduce both the static value at $\omega=0$ ($z_\omega=1$),
\begin{equation}
    \left.
    \gamma_0
    \Pi^\mathrm{D}_{{\rm sp},\perp}
    \right|_{\omega=0}
    =
    \mat{M}_1,
\end{equation}
and the leading large-$|z_\omega|$ behavior,
\begin{equation}
    \gamma_0
    \Pi^\mathrm{D}_{{\rm sp},\perp}
    =
    \frac{\mat{M}_0}{z_\omega}
    +
    \mathcal{O}(z_\omega^{-3}).
\end{equation}
For $\Delta_{\bk}\ll1$, the approximation recovers the conventional diffusion approximation in the linear-in-$\omega$ coefficient, while for $\Delta_{\bk}\gg1$ it also reproduces the leading low-frequency behavior (the errors are of the order $\mathcal{O}(\Delta_k^4)$ and $\mathcal{O}(\Delta_k^{-4})$, respectively). 
The corresponding asymptotic expansions are given in Appendix~\ref{app:low_frequency_expansion}. 
In this sense, Eq.~\eqref{eq:matrix_rational_approx} may be viewed as a matrix Pad\'e-type approximation interpolating between the static and large-$|z_\omega|$ limits.

It is worth noting that particle number conservation is preserved at all times $t$, even after introducing the matrix-valued rational approximation, when the Diffuson contribution $\Phi^\mathrm{D}$ and the Drude--Boltzmann contribution $\Phi^0$ are combined. 
This is because the density and spin sectors of the Diffuson decouple [Eq.~\eqref{eq:density_spin_blocks}], and the matrix-valued rational approximation modifies only the spin sector, which does not contribute to the total particle number. 
Since $\Gamma^\mathrm{D}_{00}(\omega) = \gamma_0/(1-\gamma_0\Pi^\mathrm{D}_{00}(\omega))=\gamma_0(1 + i/\tau\omega)$ [see Eq.~\eqref{eq:PiD_density_approx}], the density sector of $\Phi^\mathrm{D}$ contains a particle-number-conserving contribution associated with a pole at $\omega=0$, as well as a time dependent contribution. The latter exactly cancels the contribution from $\Phi^0$ after integration over momentum.

\subsection{Quadratic eigenvalue formulation of the Diffuson pole equation}
\label{sec:QEP}

Having obtained the matrix approximation for the spin sector of the Diffuson, we now determine its pole structure.
Restricting Eq.~\eqref{eq:Diffuson_pole_general} to the spin sector, the complex frequencies $\omega$ satisfy
\begin{align}
    0
    &=
    \det\!\left[
        \unit_3
        -
        \gamma_0\Pi^\mathrm{D}_{\mathrm{sp}}(E,\omega)
    \right]
    \notag\\
    &=
    \det\mat{F}(z_\omega),
    \label{eq:pole_equation}
\end{align}
with 
\begin{equation}
    \mat{F}(z)
    =
    \unit_3
    -
    \frac{\mat{C}}{z}
    -
    z
    \mat{M}_0^{1/2}
    \left(
        z^2\unit_3+\mat{L}
    \right)^{-1}
    \mat{M}_0^{1/2}.
    \label{eq:def_Fz}
\end{equation}
For the uniform SU(2) gauge fields considered in Ref.~\cite{Kakoi_2026b}, the spin sector can be decomposed into independent modes, so that the pole condition reduces to a cubic equation for each mode. 
For the general momentum-dependent SOC field considered here, such a decomposition is not available in general. 
Nevertheless, the pole condition takes the form of a rational eigenvalue problem, to which linearization techniques can be applied~\cite{Guttel_2017_review,Guttel_2014}.
Using the specific rational structure of Eq.~\eqref{eq:matrix_rational_approx}, we recast the present pole equation as a quadratic eigenvalue problem.

To find the roots of Eq.~\eqref{eq:pole_equation}, we use the equivalent condition that there exists a nonzero three-component vector $\bm{v}$ such that
\begin{equation}
    \mat{F}(z_\omega) \bm{v} = 0.
    \label{eq:pole_vector_equation}
\end{equation}
To eliminate the matrix inverse in Eq.~\eqref{eq:pole_vector_equation}, we introduce an auxiliary vector
\begin{equation}
    \bm{y}
    =
    \left(
        z_\omega^2\unit_3+\mat{L}
    \right)^{-1}
    \mat{M}_0^{1/2}\bm{v},
    \qquad
    \bm{u}
    =
    \begin{pmatrix}
        \bm{v}\\
        \bm{y}
    \end{pmatrix}.
    \label{eq:auxiliary_vector}
\end{equation}
The pole equation can then be rewritten as the six-dimensional quadratic eigenvalue problem
\begin{equation}
    \left(
        \mat{Q}_0
        +
        z_\omega\mat{Q}_1
        +
        z_\omega^2\mat{Q}_2
    \right)
    \bm{u}
    =
    0,
    \label{eq:quadratic_eigenproblem}
\end{equation}
with
\begin{equation}
\begin{gathered}
    \mat{Q}_0
    =
    \begin{pmatrix}
        -\mat{C} & 0 \\
        -\mat{M}_0^{1/2} & \mat{L}
    \end{pmatrix},
    \qquad
    \mat{Q}_1
    =
    \begin{pmatrix}
        \unit_3 & 0 \\
        0 & 0 
    \end{pmatrix},\\
    \mat{Q}_2
    =
    \begin{pmatrix}
        0 & -\mat{M}_0^{1/2}\\
        0 & \unit_3
    \end{pmatrix}.
    \label{eq:Q_matrices}
\end{gathered}
\end{equation}

The quadratic eigenvalue problem can be linearized~\cite{Tisseur_2001_review} by introducing
\begin{equation}
    \bm{w}
    =
    \begin{pmatrix}
        \bm{u}\\
        z_\omega\bm{u}
    \end{pmatrix}.
\end{equation}
Equation~\eqref{eq:quadratic_eigenproblem} is then equivalent to the $12$-dimensional generalized eigenvalue problem
\begin{equation}
    \underbrace{
    \begin{pmatrix}
        0 & \unit_6\\
        -\mat{Q}_0 & -\mat{Q}_1
    \end{pmatrix}
    }_{\mat{A}}
    \bm{w}
    =
    z_\omega
    \underbrace{
    \begin{pmatrix}
        \unit_6 & 0 \\
        0 & \mat{Q}_2
    \end{pmatrix}
    }_{\mat{B}}
    \bm{w}.
    \label{eq:generalized_eigenproblem}
\end{equation}
Since $\mat{Q}_2$ is singular by construction, not all twelve generalized eigenvalues are finite.
In the present problem, there are nine finite eigenvalues and three eigenvalues at infinity, see Appendix~\ref{app:eigenvalue_counting} for details.
The nine finite eigenvalues are consistent with the uniform SU(2) gauge-field case~\cite{Kakoi_2026b}, in which the spin sector decomposes into three modes, each governed by a cubic equation.
Thus, Eq.~\eqref{eq:generalized_eigenproblem} provides the corresponding matrix generalization for a general momentum-dependent SOC field.
We solve Eq.~\eqref{eq:generalized_eigenproblem} for the nine finite generalized eigenvalues $z_n$ and convert them to the corresponding pole frequencies through
\begin{equation}
\omega_n
=
\frac{i}{\tau}
\left(
z_n-1
\right).
\label{eq:z_to_omega}
\end{equation}
Moreover, the positive semidefiniteness of $\mat{L}$ guarantees that $\operatorname{Im}\omega_n\leq0$ for all finite poles,\footnote{\begin{minipage}[t]{\linewidth}
Let $z$ be a finite pole and $\bm{v}$ the corresponding nonzero vector satisfying Eq.~\eqref{eq:pole_vector_equation}.
Writing $x=\operatorname{Re}z$, for $x>0$ and any eigenvalue $l\geq0$ of $\mat{L}$,
\[
    \left|
        \frac{z}{z^2+l}
    \right|
    =
    \frac{1}{|z+l/z|}
    \leq
    \frac{1}{x}.
    \label{eq:bound_z}
\]
Using also $|1/z|\leq1/x$, Eq.~\eqref{eq:pole_vector_equation} gives
\[
    \|\bm{v}\|^2
    \leq
    \frac{1}{x}
    \left(
        \bm{v}^\dagger\mat{C}\bm{v}
        +
        \bm{v}^\dagger\mat{M}_0\bm{v}
    \right)
    =
    \frac{1}{x}\|\bm{v}\|^2,
\]
where $\mat{C}+\mat{M}_0=\unit_3$ is used.
Therefore, $x>1$ is impossible, and hence $\operatorname{Re}z_n\leq1$, or equivalently $\operatorname{Im}\omega_n\leq0$.
\end{minipage}}
so that the rational approximation does not generate exponentially growing spin modes.

\subsection{Pole residues and density-matrix dynamics}
\label{sec:pole_residues}

To construct the disorder-averaged density-matrix dynamics, both the pole positions and their residues are required.
Near a finite simple pole $z_n$, the spin sector of the Diffuson can be expanded as
\begin{align}
    \Gamma_{\mathrm{sp}}^\mathrm{D}(E,\omega)
    &=
    \gamma_0\mat{F}(z_\omega)^{-1}
    \notag\\
    &=
    \gamma_0
    \frac{\mat{\mathcal{R}}_n}{z_\omega-z_n}
    +
    \text{regular terms},
    \label{eq:GammaD_pole_expansion}
\end{align}
where
\begin{equation}
    \mat{\mathcal{R}}_n
    =
    \frac{
        \bm{v}_n\bm{v}_n^\top
    }{
        \bm{v}_n^\top
        \left[
            \partial_z\mat{F}(z_n)
        \right]
        \bm{v}_n
    }.
    \label{eq:Diffuson_residue}
\end{equation}
Here, $\bm{v}_n$ is a nonzero null vector satisfying $\mat{F}(z_n)\bm{v}_n=0$.
Equation~\eqref{eq:Diffuson_residue} follows from the standard residue formula for nonlinear eigenvalue problems~\cite{Keldysh_1951,*Keldysh_1971,Guttel_2017_review}.\footnote{
In general, the residue matrix of $\mat{F}(z)^{-1}$ at a simple pole involves right and left null vectors satisfying $\mat{F}(z_n)\bm{v}_n=0$ and $\bm{u}_n^\dagger\mat{F}(z_n)=0$, respectively.
Here, $\mat{C}$, $\mat{M}_0$, and $\mat{L}$ are real symmetric, so that $\mat{F}(z)^\top=\mat{F}(z)$ and the left null vector can be chosen as $\bm{u}_n=\bm{v}_n^*$.
Consequently, Eq.~\eqref{eq:Diffuson_residue} involves only $\bm{v}_n$.
Keldysh's theorem also provides the Laurent expansion at higher-order poles.
}
The derivative in the denominator is
\begin{align}
    \partial_z\mat{F}(z)
    &=
    \frac{\mat{C}}{z^2}
    -
    \mat{M}_0^{1/2}
    \left(
        z^2\unit_3+\mat{L}
    \right)^{-1}
    \mat{M}_0^{1/2}
    \notag\\
    &\quad
    +
    2z^2
    \mat{M}_0^{1/2}
    \left(
        z^2\unit_3+\mat{L}
    \right)^{-2}
    \mat{M}_0^{1/2}.
    \label{eq:F_derivative}
\end{align}
The corresponding contribution to the density matrix is obtained by substituting Eq.~\eqref{eq:GammaD_pole_expansion} into the intensity propagator $\Phi^{\mathrm{D}}$ in Eq.~\eqref{eq:PhiD_residue}, and then evaluating Eq.~\eqref{eq:density_matrix_from_Phi}.
For $\bq=\bm{0}$, the density component is conserved and decoupled from the spin sector [Eq.~\eqref{eq:density_spin_blocks}], so that the nontrivial time dependence is determined by the spin-sector $\Gamma_{\mathrm{sp}}^\mathrm{D}$.
Since $z-z_n=-i\tau(\omega-\omega_n)$, a pole $z_n$ of the Diffuson produces a contribution proportional to $e^{-i\omega_n t}$.
Consequently, $\operatorname{Re}\omega_n$ gives its oscillation frequency, $-\operatorname{Im}\omega_n$ gives its decay rate, and $\mat{\mathcal{R}}_n$ enters the corresponding mode weight.
The external disorder-averaged Green-function products in Eq.~\eqref{eq:PhiD_residue} also introduce poles in the frequency integral.
For $t>0$, the integration contour is closed in the lower half of the complex $\omega$ plane, and the full density-matrix dynamics is obtained by summing the residues of the complete integrand at the enclosed poles, as in Refs.~\cite{Kakoi_2024,Kakoi_2026b}.

\subsection{Summary of computational procedure}
\label{sec:computational_procedure}

For the calculations presented below, we construct the disorder-averaged density-matrix dynamics using the quadratic eigenvalue and pole-residue formulation developed above.
The practical procedure is summarized as follows.
\begin{enumerate}
\item
Determine $\mathcal{E}_{\pm}(\bk)$ and $\hat{\br}_{\bk}$ from the clean Hamiltonian $H_0(\bk)$.
Evaluate the dimensionless branch splitting $\Delta_{\bk}$ from Eq.~\eqref{eq:z_Delta}.

\item
Evaluate matrix $\mat{C}$ from Eq.~\eqref{eq:C_FS}.
Construct $\mat{M}_0$ and $\mat{M}_1$ from Eq.~\eqref{eq:M0M1}, and obtain $\mat{L}$ from Eq.~\eqref{eq:L_definition}.

\item
Construct the matrices $\mat{Q}_0$, $\mat{Q}_1$, and $\mat{Q}_2$ from Eq.~\eqref{eq:Q_matrices}, which define the quadratic eigenvalue problem in Eq.~\eqref{eq:quadratic_eigenproblem}.

\item 
Linearize the quadratic eigenvalue problem according to Eq.~\eqref{eq:generalized_eigenproblem} and solve 
\[
    \mat{A}\bm{w}_n = z_n\mat{B}\bm{w}_n 
\]
for the nine finite eigenvalues $z_n$.
Discard any zero eigenvalues, which do not correspond to poles of the Diffuson.

\item
For each finite eigenvalue $z_n$, extract the corresponding null vector $\bm{v}_n$ and evaluate the residue matrix $\mat{\mathcal{R}}_n$ from Eq.~\eqref{eq:Diffuson_residue}.

\item
Construct the disorder-averaged density matrix by substituting the pole expansion in Eq.~\eqref{eq:GammaD_pole_expansion} into the intensity propagator in Eq.~\eqref{eq:PhiD_residue} and evaluating Eq.~\eqref{eq:density_matrix_from_Phi}.

\end{enumerate}

\section{Applications and numerical validation}
\label{sec:demonstration}

In the previous section, we developed a general formalism for constructing the time evolution of the disorder-averaged density matrix for Hamiltonians satisfying Eq.~\eqref{eq:symmetry_condition}, using an appropriate matrix-valued rational approximation for the Diffuson.
In this section, we apply this approach to specific Hamiltonians and test its validity through direct comparison with numerical simulations.
In the numerical simulations, we consider an initially spin polarized plane-wave state and calculate the real-time evolution of the wave function using the split-step method~\cite{Weideman_1986}.
The spin expectation values $S_i(t)=\langle\psi(t)|\sigma_i|\psi(t)\rangle$ ($i=x,y,z$) are evaluated at each time and ensemble-averaged over multiple independent disorder realizations.
The analytical calculations are performed independently of the numerical simulations, following the procedure described in Sec.~\ref{sec:computational_procedure}.
The Drude--Boltzmann contribution $\Phi^0$ is also taken into account.
In Sec.~\ref{sec:channel_decoupling_and_fixed_radius}, for certain Hamiltonians, we show that further simplification for evaluating the density matrix dynamics applies.

\subsection{Two-dimensional Rashba--Dresselhaus model}
\label{sec:RD_model}

As a first application, we consider a 2D electron gas with Rashba and linear and cubic Dresselhaus SOCs~\cite{Bychkov_1984,Dresselhaus_1955}.
The clean Hamiltonian in Eq.~\eqref{eq:Hamiltonian} is given by
\begin{equation}
    \varepsilon_0(\bk)
    =
    \frac{\hbar^2\bk^2}{2m},
    \quad
    \br(\bk)
    =
    \begin{pmatrix}
        \alpha k_y-\beta k_x+\gamma k_xk_y^2\\[1mm]
        -\alpha k_x+\beta k_y-\gamma k_yk_x^2\\[1mm]
        0
    \end{pmatrix}.
    \label{eq:RD_cubic_R}
\end{equation}
Here, $\alpha$ and $\beta$ are the Rashba and linear Dresselhaus coupling constants, respectively, and $\gamma$ is the cubic Dresselhaus coupling constant.
It is useful to express $\br(\bk)$ in polar coordinates $k_x=k\cos\phi$ and $k_y=k\sin\phi$,
\begin{equation}
\begin{aligned}
    R_x(k,\phi)
    &=
    \alpha k\sin\phi
    -
    \left(
        \beta\!-\!\frac{\gamma k^2}{4}
    \right)k\cos\phi
    -
    \frac{\gamma k^3}{4}\cos3\phi,
    \\
    R_y(k,\phi)
    &=
    -\alpha k\cos\phi
    +
    \left(
        \beta\!-\!\frac{\gamma k^2}{4}
    \right)k\sin\phi
    -
    \frac{\gamma k^3}{4}\sin3\phi.
\end{aligned}
\label{eq:RD_harmonics}
\end{equation}
Thus, the cubic Dresselhaus term contains first and third angular harmonics, with the former renormalizing the linear Dresselhaus coupling~\cite{Iordanskii_1994}.
To characterize their relative strengths, we define
\begin{equation}
    \beta_3
    =
    \frac{\gamma k_0^2}{4},
    \qquad
    \widetilde{\beta}
    =
    \beta-\beta_3,
    \label{eq:effective_Dresselhaus}
\end{equation}
at a reference wave number $k_0$.\footnote{
The reference wave number enters only the definitions of $\widetilde{\beta}$ and $\beta_3$; all calculations use the full momentum-dependent field in Eq.~\eqref{eq:RD_cubic_R}.
An analogous parametrization is used, e.g., in Ref.~\cite{Wenk_2010}.
}
In this paper, $k_0$ is the initial wave number.
Here, $\widetilde{\beta}$ and $\beta_3$ are the effective first-harmonic and third-harmonic Dresselhaus coupling strengths, respectively.
For convenience, we introduce
\begin{equation}
    \kappa
    =
    \frac{m}{\hbar^2}
    \sqrt{
        2\left(
            \alpha^2+\widetilde{\beta}^{\,2}
        \right)
    },
    \qquad
    \eta
    =
    \arctan
    \left|
        \frac{
            \alpha-\widetilde{\beta}
        }{
            \alpha+\widetilde{\beta}
        }
    \right|.
    \label{eq:RD_kappa_eta}
\end{equation}
After suitable rotations of the momentum and spin axes, the first-harmonic field takes the canonical form~\cite{Kakoi_2026a}
\begin{equation}
    \br^{(1)}(\bk)
    =
    \frac{\hbar^2\kappa}{m}
    \begin{pmatrix}
        k_x'\cos\eta\\
        k_y'\sin\eta\\
        0
    \end{pmatrix},
    \label{eq:RD_linear_canonical}
\end{equation}
where $k_x'$ and $k_y'$ denote the momentum components along the rotated axes.
Thus, $\kappa$ sets the overall spin--orbit wave number scale, while $\eta$ characterizes the anisotropy of the effective linear SOC.
We further characterize the cubic coupling by the dimensionless ratio
\begin{equation}
    r_3
    =
    \frac{m\beta_3}{\hbar^2\kappa}.
    \label{eq:RD_r3}
\end{equation}
The parameter $r_3$ measures the cubic Dresselhaus coupling relative to the effective linear coupling. 
The limit $\eta=0$, corresponding to $\alpha=\widetilde{\beta}$, is the persistent spin helix (PSH) point of the effective linear SOC, where an SU(2) symmetry protects a helical spin mode~\cite{Schliemann_2003,Bernevig_2006}. 
The third-harmonic component of the cubic Dresselhaus coupling breaks the SU(2) symmetry and gives the PSH mode a finite lifetime~\cite{Stanescu_2007,Koralek_2009,Wenk_2010,Luffe_2011,Kohda_2012}.

The disordered potential is characterized by the mean free path $\ell=\overline{v}\tau$, where $\tau$ is the elastic scattering time and $\overline{v}$ is the average group velocity over the constant-energy shell.
The corresponding disorder correlation strength $\gamma_0$ is then obtained from Eq.~\eqref{eq:tau_gamma}. 
To summarize, the system is specified by the four dimensionless parameters $(k_0\ell,\kappa\ell,\eta,r_3)$. 
Here,  $\kappa\ell$ characterizes the SOC strength relative to disorder scattering. 
Unless otherwise specified, we fix $k_0\ell=40$, for which the weak disorder condition $k_0\ell\gg1$ is well satisfied.

\subsubsection{Spin dynamics: comparison with numerical simulations}

For the numerical simulations, the $\delta$-correlated disorder is represented by a spatially uncorrelated random potential on a $512\times 512$ lattice with spacing $a$, whose value at each lattice site is independently drawn from a uniform distribution over $[-\frac W2,\frac W2]$. 
The disorder amplitude $W$ is fixed by $\gamma_0=\frac{1}{12}W^2a^d$, so that the lattice disorder reproduces the continuum white-noise correlation in Eq.~\eqref{eq:disorder_correlation}~\cite{Kakoi_2026a}.
For the initial state, we take a spin polarized plane wave with momentum $\bk_0=(k_0,0)$ and spin polarization $\bm{S}(0)=\frac{1}{\sqrt{3}}(1,1,1)$.
We use a momentum cutoff $k_{\max}=2k_0$ along each Cartesian direction, corresponding to a lattice spacing of $a=\pi/k_{\max}$.
The time range considered here is much shorter than the Thouless time, so finite-size effects can be neglected.

In the analytical calculation, we resolve the initial density matrix in the branch basis because the initial spin state is generally not an eigenstate of $H_0(\bk_0)$:
\begin{equation}
    \varrho(0)
    =
    \!\sum_{s,s'=\pm} \!\!
    \varrho_{ss'}(0),
    \quad
    \varrho_{ss'}(0)
    =
    P_s(\bk_0)\varrho(0)P_{s'}(\bk_0),
    \label{eq:RD_branch_decomposition}
\end{equation}
where $P_s(\bk_0)$ projects onto branch $s$.
When the branch splitting is not negligible compared with the disorder broadening, the analytical time evolution of each diagonal component $\varrho_{ss}(0)$ is constructed using the constant-energy shell at $\mathcal{E}_s(\bk_0)$, whereas the off-diagonal components are treated at the mean branch energy $[\mathcal{E}_+(\bk_0)+\mathcal{E}_-(\bk_0)]/2$.
The resulting contributions are then summed.

\begin{figure}[t]
\centering

\includegraphics[width=\columnwidth]{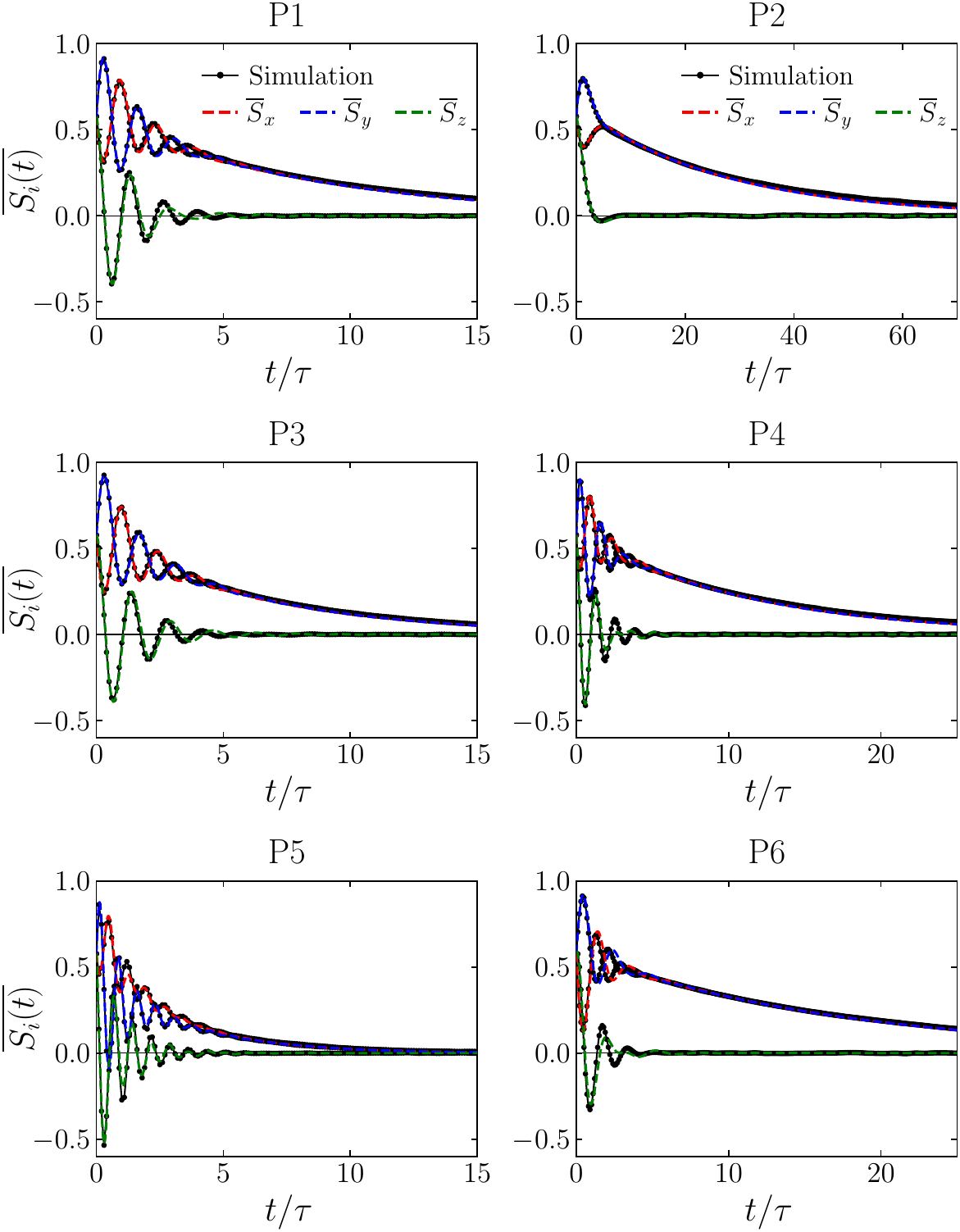}
\caption{
Comparison of the spin dynamics obtained from numerical simulations
and analytical calculations for the six representative parameter
sets P1--P6 listed in Table~\ref{tab:params_spin_dynamics}.
The initial spin polarization is
$\bm{S}(0)=\frac{1}{\sqrt{3}}(1,1,1)$, and the spin components are defined by
$\overline{S_i(t)}
=\overline{\langle\psi(t)|\sigma_i|\psi(t)\rangle}$
for $i=x,y,z$.
Black circles connected by solid lines represent the numerical
simulations, while the red, blue, and green dashed lines represent
the analytical results for $\overline{S}_x$, $\overline{S}_y$,
and $\overline{S}_z$, respectively.
}
\label{fig:spin_dynamics}

\vspace{1em}

\begin{minipage}{\columnwidth}
\centering
\captionof{table}{
Parameter sets used in Fig.~\ref{fig:spin_dynamics}.
\label{tab:params_spin_dynamics}
}
\begin{ruledtabular}
\renewcommand{\arraystretch}{1.2}
\begin{tabular}{cccc}
Name & $\kappa\ell$ & $\eta$ & $r_3$ \\
\hline
P1 & $2.5$ & $\pi/24$ & $0.3$ \\
P2 & $0.5$ & $\pi/24$ & $0.3$ \\
P3 & $2.5$ & $\pi/12$ & $0.3$ \\
P4 & $2.5$ & $0$      & $0.3$ \\
P5 & $2.5$ & $\pi/24$ & $1$   \\
P6 & $2.5$ & $\pi/24$ & $0$   \\
\end{tabular}
\end{ruledtabular}
\end{minipage}

\end{figure}

In Fig.~\ref{fig:spin_dynamics}, we compare the numerical simulations of the spin dynamics with the analytical results for the six representative parameter sets P1--P6 listed in Table~\ref{tab:params_spin_dynamics}.
For each parameter set, the numerical results are averaged over $10$ independent disorder realizations.
The analytical results agree well with the numerical simulations, reproducing the oscillatory spin relaxation that cannot be described by a simple exponential decay, although some deviations remain in the oscillation phases.
P1 serves as the reference parameter set.
It lies in a regime that is difficult to treat within conventional diffusive approaches, characterized by a spin--orbit length shorter than the mean free path ($\kappa\ell\gtrsim1$), strongly anisotropic SOC ($\eta\ll1$), and a finite cubic Dresselhaus coupling.
The comparison between P1 and P2 shows the dependence on the SOC strength $\kappa\ell$: the spin relaxation becomes slower as the SOC is reduced.
The sets P1, P3, and P4 illustrate the dependence on $\eta$.
Because of the finite cubic Dresselhaus coupling, spin relaxation persists even at the PSH condition of the linear sector, $\eta=0$, although it becomes slower.
Finally, P1, P5, and P6 show the dependence on the cubic coupling $r_3$.
The quantitative agreement across these different parameter regimes demonstrates that the rational approximation provides an accurate representation of the $\omega$ dependence of the Diffuson.

\subsubsection{Spin relaxation times}

To illustrate the utility of the matrix-valued rational approximation, we consider spin relaxation.
To place the present results in context, we first summarize the regimes covered by previous theories of Dyakonov--Perel (DP) spin relaxation~\cite{Dyakonov_1972}.
We denote the spin relaxation time by $\tau_s$ and define the average spin precession frequency as $\Omega=2\big\langle|\br(\bk)|\big\rangle_{\mathrm{FS}}/\hbar$.
In the weak SOC regime ($\Omega\tau\ll1$), the conventional DP scaling is $\tau_s\propto\Omega^{-2}\tau^{-1}$.
This $\tau^{-1}$ dependence reflects motional narrowing: frequent momentum scattering interrupts spin precession and thereby suppresses spin relaxation.
For $\Omega\tau\gtrsim1$, this scaling no longer holds, and the relaxation time typically approaches $\tau_s\sim\tau$ in the strong SOC limit ($\Omega\tau\gg1$)~\cite{Dyakonov_2008_book}.
In the Rashba limit, Burkov and Balents derived an analytic expression for the spin diffusion propagator valid for arbitrary $\Omega\tau$~\cite{Burkov_2004b}.
The poles satisfy a cubic equation that describes the crossover in spin relaxation from the weak to the strong SOC regime~\cite{Szolnoki_2017}.
Characterizing the SOC by its Fermi-surface-averaged splitting is adequate for an isotropic field such as pure Rashba SOC, but becomes insufficient when the splitting varies strongly along the Fermi surface.
A prominent example is the equal-Rashba--Dresselhaus limit associated with the PSH.
For weak SOC ($\kappa\ell\ll1$), approximate analytical expressions for the Rashba--Dresselhaus (RD) model have been obtained over the $(\kappa\ell,\eta,r_3)$ parameter space~\cite{Stanescu_2007,Wenk_2010}.
Crossover to the strong SOC regime is also investigated using other approaches~\cite{Bernevig_2008,X-Liu_2011}.
Restricting the model to $r_3=0$, Ref.~\cite{Kakoi_2026b} treated arbitrary $\kappa\ell$ and derived a Burkov--Balents-type cubic equation describing spin relaxation for any $\eta$.
The resulting spin relaxation times show quantitative agreement with numerical simulations throughout the crossover between the diffusive and ballistic regimes.
The matrix-rational formulation developed here applies to a broader class of Hamiltonians satisfying Eq.~\eqref{eq:symmetry_condition}, including the RD model with cubic couplings, as demonstrated below.

Since all finite poles determined by Eq.~\eqref{eq:generalized_eigenproblem}  satisfy $\operatorname{Im}\omega_n\leq0$, the pole closest to the real-frequency axis with a nonzero weight in $\Phi^{\rm D}$ determines the longest spin relaxation timescale for the given initial density matrix.
We define the spin relaxation time as
\begin{equation}
    \tau_{\mathrm{s}}
    =
    \frac{1}{
    \min_n|\operatorname{Im}\omega_n|
    }
    =
	\frac{\tau}{
		\min_n
		\left|
			1-\operatorname{Re}z_n
		\right|
	},
    \label{eq:spin_relaxation_time}
\end{equation}
where the minimum is taken over the poles with nonzero weights.
When the longest-lived mode satisfies $\tau_{\mathrm{s}}\gg\tau$, its pole lies in the low-frequency regime $|\omega\tau|\ll1$.
In this case, a first-order expansion of $\unit_3-\gamma_0\Pi^\mathrm{D}_{\mathrm{sp}}$ in $\omega$ reduces the pole equation to a $3\times3$ linear generalized eigenvalue problem, from which the longest-lived mode can be obtained more simply.
Details are given in Appendix~\ref{app:low_frequency}.
Note that this linear-in-$\omega$ approximation captures the long-time relaxation but does not reproduce the full short-time dynamics.

\begin{figure}[t]
\begin{center}
    \includegraphics[width=\columnwidth]{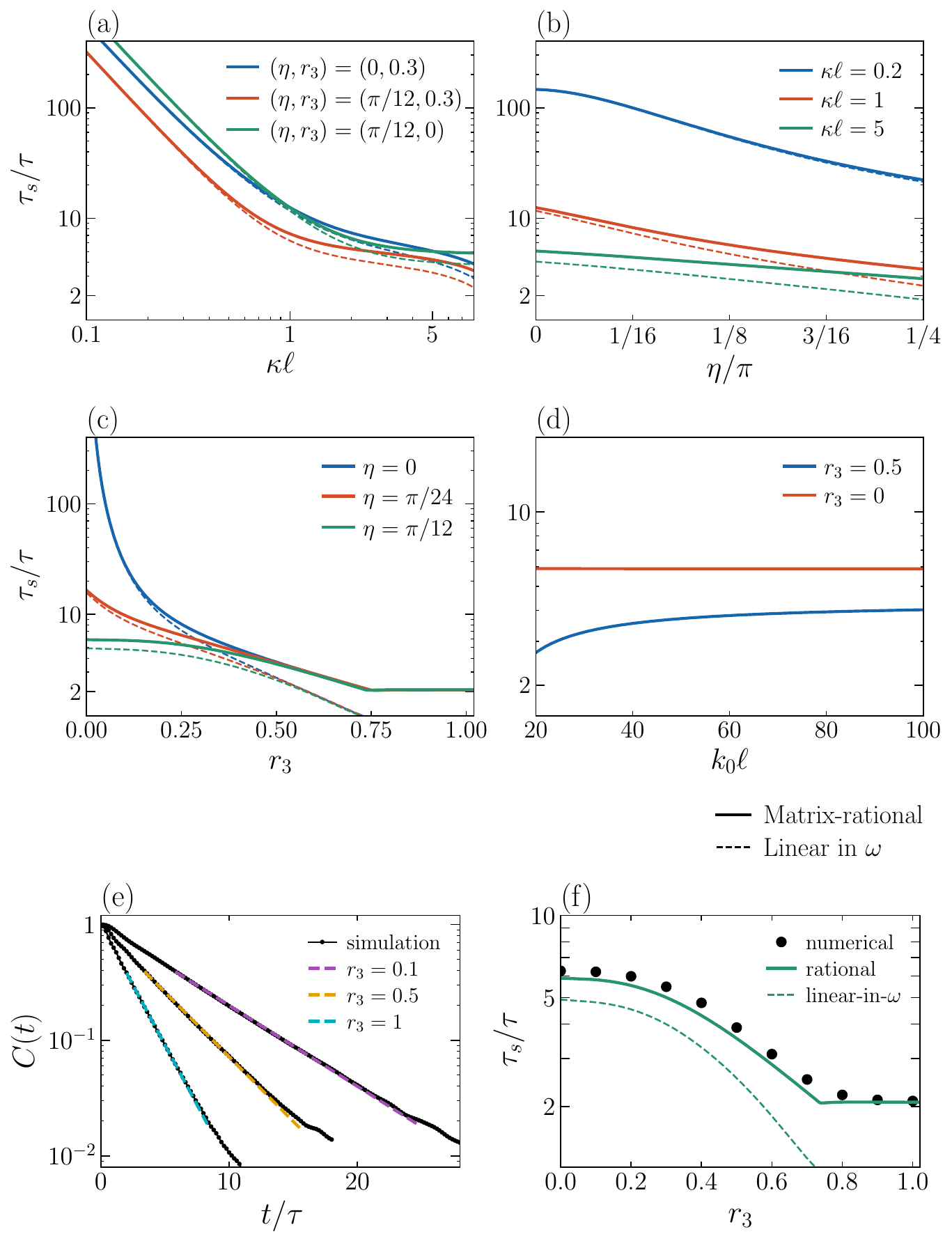}
    \caption{
    Spin relaxation times $\tau_s$ for the 2D Rashba--Dresselhaus model.
    The normalized relaxation time $\tau_s/\tau$ is shown as a function of
    (a)~the SOC strength $\kappa\ell$ for three combinations of $(\eta,r_3)$,
    (b)~the SOC anisotropy $\eta$ at $r_3=0.3$ for $\kappa\ell=0.2$, $1$, and $5$,
    (c)~the cubic Dresselhaus coupling $r_3$ at $\kappa\ell=2.5$ for $\eta=0$, $\pi/24$, and $\pi/12$, and
    (d)~the disorder parameter $k_0\ell$ at $\kappa\ell=0.5$ and $\eta=\pi/12$ for $r_3=0$ and $0.5$.
    Solid and dashed lines in panels (a)--(d) show the matrix-rational and linear-in-$\omega$ results, respectively.
    (e)~Disorder-averaged spin autocorrelation function     $C(t):=\overline{\braket{\bm{S}(t)\cdot\bm{S}(0)}}$ for $r_3=0.1$, $0.5$, and $1$ at
    $(\kappa\ell,\eta)=(2.5,\pi/12)$.
    Black circles connected by solid lines show the numerical simulations, and the colored dashed lines show exponential fits using the data satisfying $0.02<C(t)<0.4$.
    (f)~Spin relaxation times extracted from the numerical simulations (circles), compared with the analytical spin relaxation time $\tau_s/\tau$ obtained from Eq.~\eqref{eq:spin_relaxation_time} within the matrix-valued rational approximation (solid line) and the linear-in-$\omega$ approximation (dashed line).
    }
    \label{fig:spin_relaxation}
\end{center}
\end{figure}

In Figs.~\ref{fig:spin_relaxation}(a)--(d), we show the spin relaxation times obtained from the matrix-valued rational approximation and compare them with those from the linear-in-$\omega$ approximation.
As shown in Fig.~\ref{fig:spin_relaxation}(a), the weak SOC regime exhibits the known DP scaling $\tau_s/\tau\propto(\kappa\ell)^{-2}$.
When $\tau_s/\tau$ is large, the two approximations give nearly identical results.
As $\kappa\ell$ increases and $\tau_s$ becomes comparable to $\tau$, however, their discrepancy becomes more pronounced.
Figure~\ref{fig:spin_relaxation}(b) shows the $\eta$ dependence for various $\kappa\ell$ at finite cubic coupling $r_3=0.3$.
For $r_3=0$, the relaxation time diverges as $\tau_s/\tau\propto\eta^{-2}$ toward the PSH point $\eta=0$~\cite{Kakoi_2026b}.
For finite $r_3$, by contrast, $\tau_s/\tau$ remains finite at $\eta=0$.
Figure~\ref{fig:spin_relaxation}(c) shows the $r_3$ dependence for various $\eta$ at $\kappa\ell=2.5$.
At $\eta=0$ and $r_3\kappa\ell\ll1$, the relaxation time follows $\tau_s/\tau\propto r_3^{-2}$.
In Figs.~\ref{fig:spin_relaxation}(a)--(c), $k_0\ell$ is fixed at $40$.
The $k_0\ell$ dependence at $(\kappa\ell,\eta)=(0.5,\pi/12)$ is shown in Fig.~\ref{fig:spin_relaxation}(d).
For $r_3=0$, $\tau_s/\tau$ is nearly independent of $k_0\ell$ provided that $\kappa/k_0\ll1$.
A finite cubic coupling, by contrast, produces a moderate dependence on $k_0\ell$ because the cubic term more strongly deforms the constant-energy shell.
As shown in Fig.~\ref{fig:spin_relaxation}(d), however, $\tau_s/\tau$ approaches a constant when $r_3\kappa/k_0$ becomes sufficiently small, converging to the fixed-radius shell limit discussed later.

In Figs.~\ref{fig:spin_relaxation}(e) and \ref{fig:spin_relaxation}(f), we compare the spin relaxation times obtained analytically with those extracted from numerical simulations.
Since the weak SOC regime $\kappa\ell\ll1$ is well understood, we focus here on $\kappa\ell=2.5$.
We use the same parameters as for the $\eta=\pi/12$ curve in Fig.~\ref{fig:spin_relaxation}(c) and vary the cubic coupling $r_3$.
To suppress oscillatory contributions, the initial state in the simulations is chosen to be the $+$-branch eigenstate of the clean Hamiltonian $H_0(\bk_0)$.
The numerical results are averaged over $50$ disorder realizations.
Figure~\ref{fig:spin_relaxation}(e) shows the disorder-averaged spin autocorrelation function $C(t)$ for representative values of $r_3$.
The numerical relaxation times are extracted from exponential fits over the range $0.02<C(t)<0.4$.
As shown in Fig.~\ref{fig:spin_relaxation}(f), the matrix-rational results agree well with the numerical simulations over the entire range of $r_3$, although they slightly underestimate the relaxation times.
In contrast, the linear-in-$\omega$ approximation shows larger deviations in this regime.
We note that the linear-in-$\omega$ approximation itself does not assume weak SOC ($\kappa\ell\ll1$).
Indeed, it can provide accurate predictions outside the weak SOC regime when $\tau_s/\tau$ is sufficiently large, as occurs near the PSH.
This comparison shows that treating the SOC nonperturbatively is not sufficient by itself; the frequency dependence of the Diffuson must also be treated accurately.

\subsection{$d$-wave altermagnet with Rashba SOC}
\label{sec:AM_model}

As a second application, we consider a 2D $d$-wave altermagnet with an out-of-plane N\'eel vector and Rashba SOC.
The coexistence of altermagnetic spin splitting and Rashba SOC has recently been studied in connection with, for example, the RKKY interaction and finite-momentum superconductivity~\cite{Amundsen_2024,Mukasa_2025}.
Here, this model provides a physically motivated test of our formulation for a system that breaks time-reversal symmetry.
The clean Hamiltonian is
\begin{equation}
    H_0(\bk)
    =
    \frac{\hbar^2\bk^2}{2m}\unit_2
    +
    \alpha
    \left(
        k_y\sigma_x-k_x\sigma_y
    \right)
    +
    \lambda_{\mathrm{AM}}
    \left(
        k_x^2-k_y^2
    \right)
    \sigma_z,
    \label{eq:AM_Hamiltonian}
\end{equation}
where $\alpha$ is the Rashba coupling constant and $\lambda_{\mathrm{AM}}$ characterizes the $d$-wave altermagnetic spin splitting.
For $\lambda_{\mathrm{AM}}\neq0$, the time-reversal condition in Eq.~\eqref{eq:TRS_condition} is not satisfied.
Nevertheless, Eq.~\eqref{eq:symmetry_condition} is satisfied by choosing
\begin{equation}
    \mathcal{G}
    =
    \left\{
        1,C_4,C_4^2,C_4^3
    \right\},
    \label{eq:AM_symmetry_set}
\end{equation}
where $C_4$ denotes a fourfold rotation about the $z$ axis in momentum space.
Using the initial wave number $k_0$ as the reference scale, we introduce
\begin{equation}
    \kappa_{\mathrm{R}}
    =
    \frac{m\alpha}{\hbar^2},
    \qquad
    \kappa_{\mathrm{AM}}
    =
    \frac{m\lambda_{\mathrm{AM}}k_0}{\hbar^2}.
    \label{eq:AM_kappa}
\end{equation}
The model is thus characterized by the three dimensionless parameters $(k_0\ell, \kappa_{\mathrm{AM}}\ell, \kappa_{\mathrm{R}}\ell)$.
The disordered potential, initial state, and numerical procedure are the same as those used for the RD model in Sec.~\ref{sec:RD_model}.
The analytical density-matrix dynamics is constructed following Sec.~\ref{sec:computational_procedure} and compared directly with the numerical time evolution.

\begin{figure}[t]
\begin{center}
    \includegraphics[width=\columnwidth]{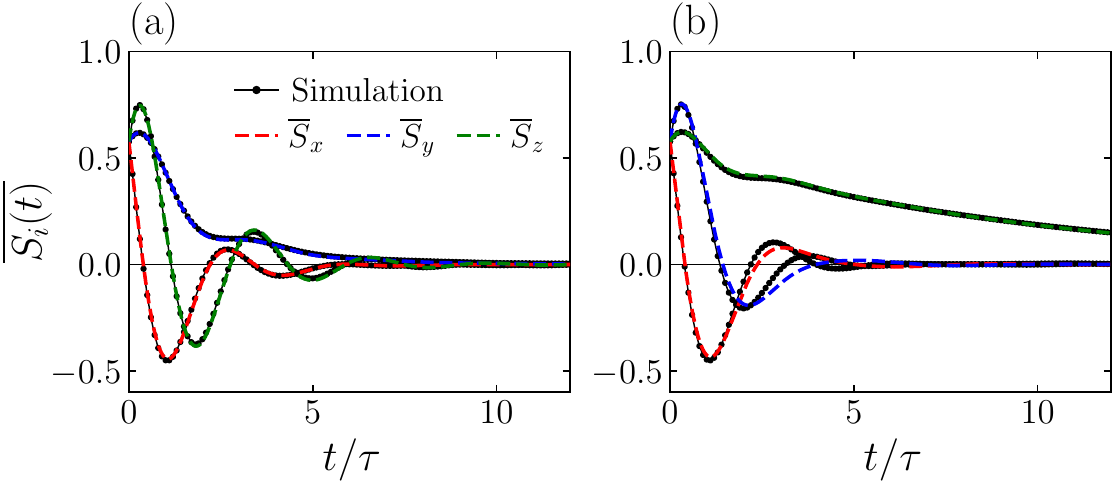}
    \caption{
    Comparison of the spin dynamics obtained from numerical simulations and analytical calculations for the $d$-wave altermagnet with Rashba SOC.
    The parameters are (a)~$(\kappa_{\mathrm{R}}\ell,\kappa_{\mathrm{AM}}\ell)=(1,\,0.25)$ and (b)~$(\kappa_{\mathrm{R}}\ell,\kappa_{\mathrm{AM}}\ell)=(0.25,\,1)$.
    }
    \label{fig:altermagnet_spin_dynamics}
\end{center}
\end{figure}

In Fig.~\ref{fig:altermagnet_spin_dynamics}, we compare the analytical results with numerical simulations in the Rashba-dominated and altermagnetism-dominated regimes.
The analytical results agree well with the numerical simulations in both cases, reproducing the damped oscillations and the distinct relaxation of the three spin components.
In particular, when the altermagnetic splitting dominates [Fig.~\ref{fig:altermagnet_spin_dynamics}(b)], the out-of-plane component relaxes substantially more slowly than the in-plane components, consistently with the conservation of $S_z$ in the limit $\kappa_\mathrm{R}=0$.

\subsection{Three-dimensional double-Weyl model}

\begin{figure}[t]
\begin{center}
    \includegraphics[width=\columnwidth]{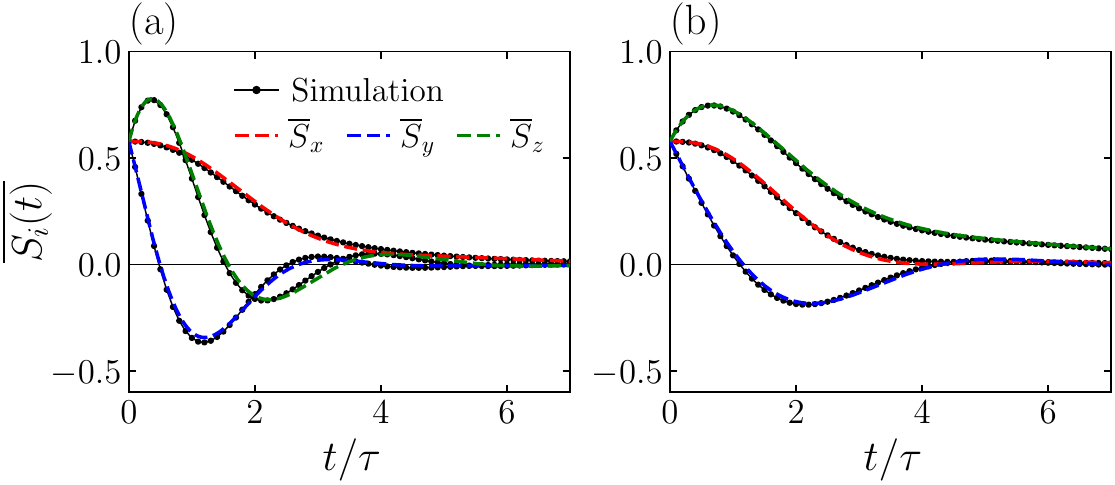}
    \caption{
    Comparison of the spin dynamics obtained from numerical simulations and analytical calculations for the 3D double-Weyl model.
    The parameters are (a)~$(\kappa_{\perp}\ell,\kappa_z\ell)=(1,\,0.5)$ and (b)~$(\kappa_{\perp}\ell,\kappa_z\ell)=(0.5,\,1)$.
    }
    \label{fig:double_weyl_spin_dynamics}
\end{center}
\end{figure}

Finally, to test the applicability of the present formulation to 3D systems with nonlinear spin--momentum coupling, we consider a two-band model with a double-Weyl point.
The clean Hamiltonian is
\begin{equation}
    H_0(\bk)
    =
    \frac{\hbar^2\bk^2}{2m}\unit_2
    +
    \lambda_{\perp}\!
    \left[
        \left(k_x^2-k_y^2\right)\sigma_x
        +
        2k_xk_y\sigma_y
    \right]
    +
    \xi_zk_z\sigma_z.
    \label{eq:double_Weyl_Hamiltonian}
\end{equation}
The pseudospin-dependent part is the standard low-energy form of a double-Weyl Hamiltonian~\cite{C-Fang_2012}.
Here, the Pauli matrices act in the effective two-band subspace (pseudospin), which we simply call spin.
Equation~\eqref{eq:symmetry_condition} is satisfied with
\begin{equation}
    \mathcal{G}
    =
    \left\{
        C_4^n,M_zC_4^n
        \,\middle|\,
        n=0,1,2,3
    \right\},
    \label{eq:double_Weyl_symmetry_set}
\end{equation}
where $M_z$ denotes the reflection across the $xy$ plane.
As in Sec.~\ref{sec:AM_model}, we parametrize the coupling strengths by
\begin{equation}
    \kappa_{\perp}
    =
    \frac{m\lambda_{\perp}k_0}{\hbar^2},
    \qquad
    \kappa_z
    =
    \frac{m\xi_z}{\hbar^2}.
    \label{eq:double_Weyl_kappa}
\end{equation}
The numerical simulations are performed on a $128\times128\times128$ spatial grid, with all other conditions chosen as in Sec.~\ref{sec:RD_model}.
As shown by the representative results in Fig.~\ref{fig:double_weyl_spin_dynamics}, the analytical predictions agree well with the numerical simulations over the parameter
range examined, demonstrating the applicability of our formulation to 3D systems with nonlinear momentum-dependent coupling.

\subsection{Further simplification through channel decoupling and the fixed-radius shell approximation}
\label{sec:channel_decoupling_and_fixed_radius}

Section~\ref{sec:computational_procedure} provides a general procedure for constructing the density-matrix dynamics of Hamiltonians satisfying Eq.~\eqref{eq:symmetry_condition}, within the weak disorder regime and the rational and on-shell approximations.
For certain Hamiltonians, the matrix structure of the spin-sector Diffuson permits a further reduction to independent scalar channels, as shown below.
Moreover, when the constant-energy shell is approximately circular or spherical, the on-shell averages reduce to angular integrals that can often be evaluated in closed form.

The spin sector of the Diffuson decouples into three independent scalar channels if $\mat{F}(z_\omega)$, defined in Eq.~\eqref{eq:def_Fz}, can be diagonalized in a $z_\omega$-independent spin basis.
Specifically, this requires an orthogonal matrix $\mat{O}$ such that
\begin{equation}
    \mat{O}^{\top}\mat{F}(z_\omega)\mat{O}
    =
    \operatorname{diag}
    \left[
        f_1(z_\omega),f_2(z_\omega),f_3(z_\omega)
    \right]
    \label{eq:decoupling_F}
\end{equation}
for all $z$.
This is possible when the real symmetric matrices $\mat{C}$, $\mat{M}_0$, and $\mat{L}$ are simultaneously diagonalizable.
Since $[\mat{C},\mat{M}_0]=0$ by definition, with $\mat{M}_0=\unit_3-\mat{C}$, the condition reduces to $[\mat{C},\mat{L}]=0$.
Using Eq.~\eqref{eq:L_definition}, this condition can equivalently be written as $[\mat{C},\mat{M}_1]=0$ on the support of $\mat{M}_0$.
Let $\be_a$ be a common eigenvector of $\mat{C}$, $\mat{M}_1$ and $\mat{L}$, with corresponding eigenvalues $c_a$, $m_a$ and $l_a$, respectively.
By definition, the relation
\begin{equation}
    l_a = \frac{1-c_a}{m_a}-1
    \label{eq:channel_la}
\end{equation}
holds.
The corresponding scalar function in Eq.~\eqref{eq:decoupling_F} is then
\begin{equation}
    f_a(z) = 1 - \frac{c_a}{z}
    - \frac{(1-c_a)z}{z^2+l_a}.
\end{equation}
The three pole equations $f_a(z_\omega)=0$ therefore reduce to the independent cubic equations
\begin{equation}\label{eq:cubic_eq}
    z_\omega^3-z_\omega^2+l_a z_\omega-c_a l_a=0.
\end{equation}
Here, $0\leq c_a\leq1$ and $l_a\geq0$.
For $c_a=0,1$, or $l_a=0$, cancellations reduce these equations to lower-order ones.
The asymptotic roots in different regimes of $c_a$ and $l_a$ are summarized in Table~\ref{tab:roots_cubic_eq}.

When the branch splitting is constant over the energy shell,
$\Delta_{\bk}=\Delta$, the matrix-rational approximation is
exact within the shell approximation, with
$\mat{L}=\Delta^2\unit_3$ and hence $l_a=\Delta^2$.
This includes pure Rashba and isotropic hedgehog SOC
within the fixed-radius shell approximation;
for the latter, $c_a=1/3$ for all three spin channels,
and Eq.~\eqref{eq:cubic_eq} reproduces the characteristic
equation of Ref.~\cite{Sumita_arXiv}.

\begin{table}[b]
    \centering
    \caption{
    Asymptotic roots $z_\omega$ of the cubic equation~\eqref{eq:cubic_eq} in different regimes of $c_a$ and $l_a$ ($c_a\neq0,1$ and $l_a\neq0$). 
    Real roots correspond to non-oscillatory modes, while complex-conjugate roots describe damped oscillations.
    }
    \begin{ruledtabular}
    \renewcommand{\arraystretch}{2}
    \begin{tabular}{ccc}
    \\[-9mm]
    Condition & Non-oscillatory & Oscillatory \\
    \hline
    $l_a\ll c_a(<1)$ & $1-(1-c_a)l_a$ & $\displaystyle \frac{(1-c_a)l_a}{2}\pm i\sqrt{c_al_a}$\\
    $c_a\ll l_a\ll1$ & $1-l_a$, $l_a$, $c_a$ & $-$\\
    $l_a\gg1$ & $c_a$ & $\displaystyle \frac{1-c_a}{2} \pm i\sqrt{l_a}$ \\
    \end{tabular}
    \end{ruledtabular}
    \label{tab:roots_cubic_eq}
\end{table}

For all three models considered in Sec.~\ref{sec:demonstration}, $\mat C$ and $\mat M_1$ commute, and the above simplification therefore applies.\footnote{
For example, adding a term $\propto k_x\sigma_z$ to the RD model ($r_3\neq0$) gives $[\mat{C},\mat{M}_1]\neq0$.
Such out-of-plane SOC terms occur in asymmetric quantum wells, e.g., $(110)$-grown structures~\cite{Poshakinskiy_2013,Iizasa_2018}.
The matrix rational approximation also describes the spin dynamics in this case, as shown in Appendix~\ref{app:noncommuting_model}.
}
For the RD model, the common eigenbasis is $\{\be_+,\be_-,\be_z\}$, where $\be_\pm=(\be_x\pm\be_y)/\sqrt{2}$.
The $d$-wave altermagnet with Rashba SOC (AM+RSOC) and the double-Weyl model are diagonal in the Cartesian basis $\{\be_x,\be_y,\be_z\}$, with degenerate $x$ and $y$ channels.
Here, the vectors $\be_a$ denote basis vectors in pseudospin space rather than directions in real space.

\begin{table*}[t]
\caption{\label{tab:channel_coefficients}
Summary of common eigenvectors and corresponding eigenvalues of $\mat{C}$ and $\mat{M}_1$ within the fixed-radius shell approximation.
The model parameters are given in parentheses below each model name.
Within this approximation, no $k_0\ell$ dependence appears.
The relevant functions are defined below the dashed lines for each model.
For the Rashba--Dresselhaus model, $\be_\pm=(\be_x\pm\be_y)/\sqrt{2}$. Note that $m_\pm$ involves $\chi^{}_\mp$.
Detailed derivations of $c_a$ and $m_a$ are provided in the Supplemental Material~\cite{SM}.
}
\begin{ruledtabular}
\renewcommand{\arraystretch}{2.4}
\begin{tabular}{
    @{}c@{\hspace{4mm}}
    c@{\hspace{10mm}}
    c@{\hspace{-2mm}}
    c@{\hspace{-10mm}}
}
\\[-12mm]
Model
    & $\be_a$ & $c_a$ & $m_a$
\\[-1mm]
\hline

\multirow{2}{*}{
    \begin{tabular}{c}
        Rashba--Dresselhaus\\[-2mm]
        ($\kappa\ell$, $\eta$, $r_3$)
    \end{tabular}
}
& $\be_z$ & $c_z=0$ 
&
$\displaystyle
m_z=\frac{1}{\pi}
\int_0^1\frac{d\mu}{\sqrt{\mu(1-\mu)}}
\frac{1}{1+4(\kappa\ell)^2\chi(\mu)}
$
\\

& $\be_{+},\,\be_{-}$
&
$\displaystyle
c_{\pm}
=
\frac{1}{\pi}
\int_0^1
\frac{d\mu}{\sqrt{\mu(1-\mu)}}
\frac{\chi_{\pm}(\mu)}{\chi(\mu)}
$
&
$\displaystyle
m_{\pm}
=
\frac{1}{\pi}
\int_0^1
\frac{d\mu}{\sqrt{\mu(1-\mu)}}
\frac{\chi_{\mp}(\mu)}
{\chi(\mu)\left[1+4(\kappa\ell)^2\chi(\mu)\right]}
$
\\[-4mm]

&
\multicolumn{3}{@{}l@{}}{
    \xleaders
    \hbox to 3mm{\rule[0ex]{1.5mm}{0.35pt}\hfil}
    \hfill\kern0pt
}
\\[-4mm]

&
\multicolumn{3}{l}{
$\displaystyle
\chi_+(\mu)
=
\mu\left[\cos\eta+r_3(3-4\mu)\right]^2,
\quad
\chi_-(\mu)
=
(1-\mu)\left[\sin\eta+r_3(1-4\mu)\right]^2,
\quad
\chi(\mu)=\chi_+(\mu)+\chi_-(\mu).
$
}
\\
\hline

\multirow{2}{*}{
    \begin{tabular}{c}
        $d$-wave altermagnet\\[-2mm]
        with Rashba SOC\\[-2mm]
        ($\kappa_{\mathrm{AM}}\ell$, $\kappa_{\mathrm R}\ell$)
    \end{tabular}
}
& $\be_z$
&
$\displaystyle
c_z=1-\frac{\kappa_{\mathrm R}\ell}{u}
$
&
$\displaystyle
m_z=\frac{\kappa_{\mathrm R}\ell}{u}
-4(\kappa_{\mathrm R}\ell)^2J
$
\\

& $\be_x,\,\be_y$
&
$\displaystyle
c_x=c_y=\frac{\kappa_{\mathrm R}\ell}{2u}
$
&
$\displaystyle
m_x=m_y=J-\frac{m_z}{2}
$
\\[-4mm]

&
\multicolumn{3}{@{}l@{}}{
    \xleaders
    \hbox to 3mm{\rule[0ex]{1.5mm}{0.35pt}\hfil}
    \hfill\kern0pt
}
\\[-3.5mm]

&
\multicolumn{3}{l}{
$\displaystyle
u=\sqrt{
    (\kappa_{\mathrm R}\ell)^2
    +
    (\kappa_{\mathrm{AM}}\ell)^2
},
\qquad
J^{-1}=\sqrt{
    \left[1+4(\kappa_{\mathrm R}\ell)^2\right]
    \left(1+4u^2\right)
}.
$
}
\\
\hline

\multirow{2}{*}{
    \begin{tabular}{c}
        3D double-Weyl\\[-2mm]
        ($\kappa_\perp\ell$, $\kappa_z\ell$)
    \end{tabular}
}
& $\be_z$
&
$\displaystyle
c_z=\int_0^1 d\mu\,
\frac{B(\mu)}{D(\mu)}
$
&
$\displaystyle
m_z=\int_0^1 d\mu\,
\frac{A(\mu)}
{D(\mu)\left[1+D(\mu)\right]}
$
\\

& $\be_x,\,\be_y$
&
$\displaystyle
c_x=c_y
=
\frac{1}{2}
\int_0^1 d\mu\,
\frac{A(\mu)}{D(\mu)}
$
&
$\displaystyle
m_x=m_y
=
\int_0^1 d\mu\,
\frac{1}{1+D(\mu)}
-
\frac{m_z}{2}
$
\\[-4mm]

&
\multicolumn{3}{@{}l@{}}{
    \xleaders
    \hbox to 3mm{\rule[0ex]{1.5mm}{0.35pt}\hfil}
    \hfill\kern0pt
}
\\[-4mm]

&
\multicolumn{3}{l}{
$\displaystyle
A(\mu)=4(\kappa_\perp\ell)^2(1-\mu^2)^2,
\qquad
B(\mu)=4(\kappa_z\ell)^2\mu^2,
\qquad
D(\mu)=A(\mu)+B(\mu).
$
}
\\[-1.5mm]

\end{tabular}
\end{ruledtabular}
\end{table*}

The coefficients in Eq.~\eqref{eq:cubic_eq} take a simple form under an approximation in which the radial dependence of $\br(\bk)$ is evaluated at $k=k_0$, while its full angular dependence is retained.
We refer to this as the fixed-radius shell approximation.
This approximation is justified when $|R(\bk_{\mathrm{on\text{-}shell}})|\ll\varepsilon_0(\bk_0)$, so that the deformation of the constant-energy shell due to the spin--momentum coupling is negligible (while the spin-dependent energy splitting must be retained).
For the models considered here, this condition corresponds to $\kappa/k_0,r_3\kappa/k_0\ll1$ for the RD model, $\kappa_{\mathrm{R}}/k_0,\kappa_{\mathrm{AM}}/k_0\ll1$ for the AM+RSOC model, and $\kappa_\perp/k_0,\kappa_z/k_0\ll1$ for the double-Weyl model.
Note that these conditions are compatible with the strong SOC regime $\kappa\ell\gtrsim1$, since $\kappa/k_0\ll1$ and $\kappa\ell\gtrsim1$ can hold simultaneously when $k_0\ell\gg1$.
Within this approximation, $c_a$ and $m_a$ are summarized in Table~\ref{tab:channel_coefficients}, with detailed derivations provided in the Supplemental Material~\cite{SM}.
The corresponding $l_a$ follows from Eq.~\eqref{eq:channel_la}.
These dimensionless eigenvalues and the resulting pole equations are independent of $k_0\ell$ within the fixed-radius shell approximation.
All angular integrals in Table~\ref{tab:channel_coefficients} can in fact be evaluated in terms of elementary functions.
Thus, within this approximation, the disorder-averaged density matrix of the three models can be written in closed form, although the explicit expressions are too lengthy to present.

As a simple illustration, we examine the asymptotic behavior of the spin relaxation rates in several limiting regimes of the RD model by expanding the integrands in Table~\ref{tab:channel_coefficients} and performing the angular integrations.
We first look at the weak SOC limit.
Defining the normalized spin relaxation rates as $\Gamma_{s,a}=\tau/\tau_{s,a}$, we obtain
\begin{equation}\label{eq:relaxation_weak_SOC_limit}
\begin{aligned}
    \Gamma_{s,+}
    &\simeq
    2(\kappa\ell)^2
    \left(\sin^2\eta+r_3^2\right),\\
    \Gamma_{s,-}
    &\simeq
    2(\kappa\ell)^2
    \left(\cos^2\eta+r_3^2\right),\\
    \Gamma_{s,z}
    &\simeq
    2(\kappa\ell)^2
    \left(1+2r_3^2\right),
\end{aligned}
\qquad
\text{for $\kappa\ell\ll1$}.
\end{equation}
These expressions reproduce the conventional DP scaling $\Gamma_{s,a}\propto(\kappa\ell)^2$.
Near the exact PSH point $\eta=r_3=0$, in-plane relaxation rate also reproduces the known dependences $\Gamma_{s,+}\propto\eta^2$ and $\Gamma_{s,+}\propto r_3^2$~\cite{Stanescu_2007}.
Even for $\kappa\ell\gg1$, the slowest in-plane relaxation mode can be described by a simple expression sufficiently close to the PSH point.
For $1\ll\kappa\ell\ll\eta^{-1},|r_3|^{-1}$, we find
\begin{equation}\label{eq:relaxation_PSH_limit}
    \Gamma_{s,+}
    \simeq
    2\kappa\ell(\eta+r_3)^2.
\end{equation}
Comparing Eqs.~\eqref{eq:relaxation_weak_SOC_limit} and \eqref{eq:relaxation_PSH_limit}, we find that both exhibit the scalings $\Gamma_{s,+}\propto\eta^2$ and $\propto r_3^2$ near $\eta=r_3=0$, whereas in the regime $\kappa\ell\gg1$, the scaling with $\kappa\ell$ changes from the conventional quadratic dependence to a linear one.
Equation~\eqref{eq:relaxation_PSH_limit} also contains a term proportional to $\eta r_3$.
Note that these results for the longest-lived relaxation mode, which satisfies $\Gamma_{s,+}\ll1$, are not specific to the matrix-rational approximation, as the linear-in-$\omega$ approximation gives the same results.
In this sense, the regime $\kappa\ell\sin\eta\gg1$ may be more interesting.
For $\kappa\ell|r_3|\ll1$, expanding to first order in $r_3$ yields
\begin{equation}\label{eq:relaxation_strong_SOC_limit}
    \Gamma_{s,+}
    \simeq
    \frac{1}{1+\cot\eta}
    +
    r_3\frac{1-3\sin2\eta}
    {(1+\sin2\eta)^{3/2}}.
\end{equation}
The first term was obtained in Ref.~\cite{Kakoi_2026b} and verified numerically there.
Interestingly, the sign of the linear correction in $r_3$ depends on $\eta$.
We emphasize that the nonmonotonic behavior of $\tau_s/\tau$ at large $\kappa\ell$ or $r_3$, observed in Figs.~\ref{fig:spin_relaxation}(a) and \ref{fig:spin_relaxation}(c), arises when the exact deformed constant-energy shell is used instead of the fixed-radius shell approximation.
Indeed, decreasing $\kappa/k_0$ (i.e., increasing $k_0\ell$ at fixed $\kappa\ell$) extends the range of $\kappa\ell$ or $r_3$ over which the relaxation time remains monotonic.
This illustrates the limitations of the fixed-radius shell approximation at finite $\kappa/k_0$.

\section{Discussion}

We developed a theoretical framework for describing the disorder-averaged density-matrix dynamics of generic pseudospin--momentum coupled systems. 
By introducing a matrix-valued rational approximation for the frequency dependence of the Diffuson [Eq.~\eqref{eq:matrix_rational_approx}], we reduced the pole equation to a finite-dimensional quadratic eigenvalue problem [Eq.~\eqref{eq:quadratic_eigenproblem}] and constructed the real-time evolution of the density matrix from the resulting poles and residues. 
The formulation applies to a broad class of Hamiltonians satisfying Eq.~\eqref{eq:symmetry_condition}, without restricting the momentum dependence of the coupling or its strength relative to disorder scattering. 
Unlike conventional low-frequency approaches, the present formulation captures both short-time and long-time dynamics, as demonstrated by comparisons with numerical simulations.
A noteworthy, less obvious property of the present formulation is that the positive semidefiniteness of $\mat{L}$ prevents unphysical exponentially growing modes within the rational approximation.

The post-quench spin dynamics studied here is closely related to the spin-dependent relaxation of optically excited carriers probed in pump--probe experiments~\cite{Kikkawa_1998,Wu_2010_review}.
Optical excitation generally produces a distribution over momentum and energy~\cite{J-Xu_2021}, which, in a simple idealization, can be modeled as a spin polarized population distributed over an energy shell~\cite{Pezzoli_2013,Wu_2010_review}.
Although we used an initial plane wave as a simple benchmark for the microscopic theory, the evolution of such a momentum-diagonal density matrix can be obtained in the same way by integrating over the initial distribution.
A quantitative comparison with Kerr or Faraday rotation would additionally require modeling the optical excitation and the probe response that relates the evolving density matrix to the measured signal~\cite{J-Xu_2021,Eskandari-asl_2026}.

Ultracold atoms offer a more direct way to probe these dynamics.
Time-of-flight measurements provide access to time-dependent momentum distributions~\cite{Jendrzejewski_2012,Labeyrie_2012}, and internal-state-resolved detection probes pseudospin dynamics~\cite{Hasan_2022}.
Although we have presented momentum-integrated spin expectation values, the formulation also gives the momentum-resolved density matrix and hence the spin-resolved momentum distributions~\cite{Kakoi_2024,Kakoi_2026b}.

In this paper, we have focused on the dynamics described by the Diffuson at $\bq=\bm{0}$.
The formulation can still be applied to $q\ell\lesssim1$, with only minor modifications, since the density--spin separation in Eq.~\eqref{eq:density_spin_blocks} remains approximately valid~\cite{Kakoi_2026b}.
The same framework can also be applied to the Cooperon by replacing the particle--hole propagator $\Pi^\mathrm{D}$ with the particle--particle propagator $\Pi^\mathrm{C}$.
For the models considered here, the Cooperon likewise has a singlet--triplet block structure at $\bq=\bm{0}$~\cite{Hikami_1980,Hijano_2024}.
This extension would be relevant to time-domain probes of weak (anti)localization, including the proposed current echo induced by terahertz pulses~\cite{ZL-Li_2024}.
Applications beyond this small-$q$ regime would require a more general treatment beyond a low-order expansion~\cite{Ilic_2020,Hijano_2024,Kakoi_2026a}.

Extending the present formulation beyond spin-independent, spatially $\delta$-correlated disorder remains a task for future work.
Dopant inhomogeneity can generate spatially fluctuating Rashba coupling even in symmetrically doped quantum wells~\cite{Sherman_2003}.
The resulting spin relaxation depends on the spatial correlations of the random coupling~\cite{Dugaev_2009,Glazov_2010_review}.
Spatial correlations are also relevant to scalar disorder, notably optical speckle potentials in ultracold atoms~\cite{Kuhn_2007,Richard_2019,Delande_2014,Orso_2017}.

\begin{acknowledgments}
    This work was supported by JSPS KAKENHI, Grant No.~JP25KJ1758. 
    M.K thanks Keith Slevin for carefully reading the manuscript and for helpful comments.

    The author used OpenAI ChatGPT (GPT-5.6 Sol and GPT-6.1 Sol) to assist with analytical calculations, manuscript preparation, and code review.
    After the author formulated Eqs.~\eqref{eq:pole_equation}--\eqref{eq:pole_vector_equation}, ChatGPT suggested introducing an auxiliary vector in Eq.~\eqref{eq:auxiliary_vector} to obtain the equivalent quadratic eigenvalue problem in Eq.~\eqref{eq:quadratic_eigenproblem}.
    ChatGPT also assisted with the linear-algebraic proofs in Appendix~\ref{app:eigenvalue_counting} and the footnotes, and the derivation of the coefficients in Table~\ref{tab:channel_coefficients}.
    The author takes full responsibility for this work.
\end{acknowledgments}

\appendix
\section{Derivation of $\Pi^\mathrm{D}$ in the Pauli basis}
\label{app:diffuson_derivation}

Here we derive the Pauli-basis representation of the Diffuson used in Eqs.~\eqref{eq:PiD_density}--\eqref{eq:PiD_spin_perp}.
Substituting the Green's function in Eq.~\eqref{eq:G_projector} into Eq.~\eqref{eq:PiD_chsp_trace}, we obtain
\begin{align}
    \Pi^\mathrm{D}_{ij}(\bq,E,\omega)
    &=
    \sum_{s,s'=\pm}
    \int_{\bk}
    g_s(\bk_+,E_+)
    g_{s'}^*(\bk_-,E_-)
    \notag\\
    &\qquad\times
    \Tr\!\left[
        T_i P_s(\bk_+)
        T_j P_{s'}(\bk_-)
    \right].
    \label{eq:PiD_projector_expansion}
\end{align}
We now set $\bq=\bm{0}$, as in the main text.
Using $P_s(\bk)$ in Eq.~\eqref{eq:projector} together with the Pauli-matrix algebra
\begin{equation}
\begin{aligned}
    \sigma_i\sigma_j
    &=
    \delta_{ij}\unit_2
    +
    i\epsilon_{ijk}\sigma_k,
    \\
    \Tr[\sigma_i\sigma_j]
    &=
    2\delta_{ij},
    \\
    \Tr[\sigma_i\sigma_a\sigma_j\sigma_b]
    &=
    2\left(
        \delta_{ia}\delta_{jb}
        -
        \delta_{ij}\delta_{ab}
        +
        \delta_{ib}\delta_{aj}
    \right),
\end{aligned}
\label{eq:Pauli_identities}
\end{equation}
the required traces in the Pauli basis of Eq.~\eqref{eq:Pauli_basis} are
\begin{align}
    \Tr[T_0P_sT_0P_{s'}]
    &=
    \frac{1}{4}(1+ss'),
    \label{eq:trace_density}
    \\
    \Tr[T_0P_sT_iP_{s'}]
    &=
    \frac{1}{4}(s+s')\hat{R}_{\bk,i},
    \label{eq:trace_density_spin}
    \\
    \Tr[T_iP_sT_jP_{s'}]
    &=
    \frac{1}{4}
    \left[
        (1-ss')\delta_{ij}
        +
        2ss'\hat{R}_{\bk,i}\hat{R}_{\bk,j} 
    \right. \notag\\
    &\qquad + \left.
        i(s-s')\epsilon_{iaj}\hat{R}_{\bk,a}
    \right].
    \label{eq:trace_spin}
\end{align}
Under the symmetry condition in Eq.~\eqref{eq:symmetry_condition}, the Green-function factors are identical at symmetry-related momenta, while the corresponding spin--orbit fields sum to zero. 
Hence, all terms linear in $\hat{\br}_{\bk}$ vanish after momentum integration. 
This gives $\Pi^\mathrm{D}_{0i}=\Pi^\mathrm{D}_{i0}=0$ in Eq.~\eqref{eq:trace_density_spin} and removes the last term in Eq.~\eqref{eq:trace_spin}.
For the remaining terms, the same-branch contribution $s'=s$ gives
\begin{equation}
    \Tr[T_iP_sT_jP_s]
    =
    \frac{1}{2}
    \hat{R}_{\bk,i}\hat{R}_{\bk,j},
    \label{eq:trace_spin_parallel}
\end{equation}
whereas the different-branch contribution $s'=-s$ gives
\begin{equation}
    \Tr[T_iP_sT_jP_{-s}]
    =
    \frac{1}{2}
    \left(
        \delta_{ij}
        -
        \hat{R}_{\bk,i}\hat{R}_{\bk,j}
    \right),
    \label{eq:trace_spin_perp}
\end{equation}
after the odd term has been omitted.
Substituting Eqs.~\eqref{eq:trace_density}, \eqref{eq:trace_spin_parallel}, and \eqref{eq:trace_spin_perp} into Eq.~\eqref{eq:PiD_projector_expansion}, we obtain Eqs.~\eqref{eq:PiD_density}--\eqref{eq:PiD_spin_perp}.

\section{Low-frequency expansion and its consequences}
\label{app:low_frequency_expansion}

\subsection{Consistency of the matrix rational approximation}
\label{app:matrix_approximation}

We first examine the low-frequency behavior of the matrix rational approximation.
Expanding the different-branch contribution in Eq.~\eqref{eq:PiD_spin_perp} around $\omega=0$ within the on-shell approximation gives
\begin{equation}
    \gamma_0
    \Pi^\mathrm{D}_{{\rm sp},\perp}
    =
    \mat{M}_1
    -
    i\omega\tau
    \left(
        \mat{M}_1-2\mat{M}_2
    \right)
    +
    \mathcal{O}\!\left(\omega^2\tau^2\right),
    \label{eq:low_freq_exact_appendix}
\end{equation}
where
\begin{equation}
    \mat{M}_2
    =
    \left\langle
        \frac{
            \mat{P}_{\bk}
        }{
            (1+\Delta_{\bk}^2)^2
        }
    \right\rangle_{\! E_0}
    \label{eq:M2_definition}
\end{equation}
and
\begin{equation}
    \mat{P}_{\bk} = \unit_3 - \hat{\br}_{\bk}\hat{\br}_{\bk}^\top.
\end{equation}
The matrices $\mat{M}_0$ and $\mat{M}_1$ are defined in Eq.~\eqref{eq:M0M1}.
On the other hand, Eq.~\eqref{eq:matrix_rational_approx} yields
\begin{equation}
    \gamma_0
    \Pi^\mathrm{D}_{{\rm sp},\perp}
    \simeq
    \mat{M}_1
    -
    i\omega\tau
    \left(
        \mat{M}_1
        -
        2\mat{M}_1\mat{M}_0^{-1}\mat{M}_1
    \right)
    +
    \mathcal{O}\!\left(\omega^2\tau^2\right).
    \label{eq:low_freq_approx_appendix}
\end{equation}
Thus, the difference in the coefficient of the linear-in-$\omega$ term is
\begin{equation}
    \delta\mat{K}
    =
    2
    \left(
        \mat{M}_2
        -
        \mat{M}_1\mat{M}_0^{-1}\mat{M}_1
    \right).
    \label{eq:linear_error}
\end{equation}
The difference vanishes when $\Delta_{\bk}$ is constant over the energy shell, since in this case $\mat M_1=(1+\Delta^2)^{-1}\mat M_0$ and $\mat M_2=(1+\Delta^2)^{-2}\mat M_0$.

For $\Delta_{\bk}\ll1$, define
\begin{equation}
    \mat{D}_2
    =
    \left\langle
        \Delta_{\bk}^2\mat{P}_{\bk}
    \right\rangle_{E_0}.
\end{equation}
Then
\begin{align}
    \mat{M}_1
    &=
    \mat{M}_0-\mat{D}_2
    +\mathcal{O}(\Delta_{\bk}^4),
    \\
    \mat{M}_2
    &=
    \mat{M}_0-2\mat{D}_2
    +\mathcal{O}(\Delta_{\bk}^4),
\end{align}
and 
\begin{equation}
    \mat{M}_1\mat{M}_0^{-1}\mat{M}_1 
    = \mat{M}_0-2\mat{D}_2
    +\mathcal{O}(\Delta_{\bk}^4),
\end{equation}
and therefore
\begin{equation}
    \delta\mat{K}
    =
    \mathcal{O}(\Delta_{\bk}^4).
    \label{eq:weak_SOC_error}
\end{equation}
Thus, Eq.~\eqref{eq:matrix_rational_approx} recovers the conventional diffusion approximation up to corrections of order $\mathcal{O}(\Delta_{\bk}^4)$ in the linear-in-$\omega$ coefficient.
In the opposite limit, $\Delta_{\bk}\gg1$ over the relevant energy shell, we have
\begin{align}
    \mat{M}_1
    &=
    \left\langle
        \frac{\mat{P}_{\bk}}{\Delta_{\bk}^2}
    \right\rangle_{\! E_0}
    +
    \mathcal{O}(\Delta_{\bk}^{-4}),
    \\
    \mat{M}_2
    &=
    \mathcal{O}(\Delta_{\bk}^{-4}),
\end{align}
and 
\begin{equation}
    \mat{M}_1\mat{M}_0^{-1}\mat{M}_1 
    = \mathcal{O}(\Delta_{\bk}^{-4}).
\end{equation}
It follows from Eq.~\eqref{eq:linear_error} that
\begin{equation}
    \delta\mat{K}
    =
    \mathcal{O}(\Delta_{\bk}^{-4}),
    \label{eq:strong_SOC_error}
\end{equation}
Thus, the linear-in-$\omega$ coefficient is reproduced to leading order, with corrections of order $\mathcal{O}(\Delta_{\bk}^{-4})$.
Equations~\eqref{eq:weak_SOC_error} and \eqref{eq:strong_SOC_error} show that, despite being fixed only by the static and large-$|z_\omega|$ limits, the approximation also yields the correct low-frequency asymptotics in both the small- and large-$\Delta_{\bk}$ regimes.

\subsection{Longest-lived spin relaxation mode}
\label{app:low_frequency}

The same low-frequency expansion can be used to determine the longest-lived spin mode directly.
When $\tau_s\gg\tau$, the corresponding pole satisfies $\omega\tau\ll1$.
Defining $\lambda=i\omega\tau$ and combining Eq.~\eqref{eq:low_freq_exact_appendix} with the same-branch contribution in Eq.~\eqref{eq:PiD_spin_parallel}, the pole equation to first order in frequency reduces to
\begin{equation}
	\mat{F}_0\bm{v}
	=
	\lambda\mat{F}_1\bm{v},
	\label{eq:lowfreq_GEP}
\end{equation}
where
\begin{equation}
	\mat{F}_0
	=
	\mat{M}_0-\mat{M}_1,
	\qquad
	\mat{F}_1
	=
	\mat{C}-\mat{M}_1+2\mat{M}_2.
	\label{eq:lowfreq_F01}
\end{equation}
Thus, the low-frequency pole problem reduces to a $3\times3$
generalized eigenvalue problem.
Denoting its generalized eigenvalues by $\lambda_n$, we have
$\operatorname{Im}\omega_n=-\operatorname{Re}\lambda_n/\tau$.
Equation~\eqref{eq:spin_relaxation_time} therefore gives
\begin{equation}
	\frac{\tau_s}{\tau}
	=
	\frac{1}{
		\min_n |\operatorname{Re}\lambda_n|
	}.
	\label{eq:tau_s_lowfreq}
\end{equation}

\section{Eigenvalue counting for the quadratic eigenvalue problem}
\label{app:eigenvalue_counting}

We show that the quadratic eigenvalue problem in Eq.~\eqref{eq:quadratic_eigenproblem} has nine finite eigenvalues and three eigenvalues at infinity, counted with algebraic multiplicity.
For notational simplicity, we write
\begin{equation}
    z=z_\omega,
    \qquad
    \mat{S}=\mat{M}_0^{1/2},
    \qquad
    \mat{T}(z) = z^2\unit_3+\mat{L}.
\end{equation}
The quadratic matrix polynomial can then be written as
\begin{align}
    \mat{Q}(z)
    = \mat{Q}_0 + z \mat{Q}_1 + z^2\mat{Q}_2 =
    \begin{pmatrix}
        z\unit_3-\mat{C}
        &
        -z^2\mat{S}
        \\
        -\mat{S}
        &
        \mat{T}(z)
    \end{pmatrix}.
    \label{eq:app_Q_matrix}
\end{align}
To determine the degree of $\det\mat{Q}(z)$, we consider its large-$|z|$ behavior. For sufficiently large $|z|$, the lower-right block $\mat{T}(z)$ is invertible.
Using the Schur complement, we obtain
\begin{align}
    \det\mat{Q}(z)
    &=
    \det\!\left[
        \mat{T}(z)
    \right]
    \det\!\left[
        z\unit_3-\mat{C}
        -
        z^2\mat{S}
        \mat{T}(z)^{-1}
        \mat{S}
    \right].
    \label{eq:app_detQ}
\end{align}
Since
\begin{equation}
    \mat{T}(z)^{-1}
    =
    \frac{1}{z^2}\unit_3
    +
    \mathcal{O}(z^{-4}),
\end{equation}
the large-$|z|$ behavior is
\begin{align}
    \det\mat{Q}(z)
    &=
    z^6
    \det\!\left[
        z\unit_3-\mat{C}-\mat{M}_0
        +
        \mathcal{O}(z^{-2})
    \right]
    \notag\\
    &=
    z^9
    \left[
        1+\mathcal{O}(z^{-1})
    \right].
    \label{eq:app_detQ_asymptotic}
\end{align}
Thus,
\begin{equation}
    \deg\det\mat{Q}(z)=9,
\end{equation}
and the quadratic eigenvalue problem has nine finite eigenvalues, counted with algebraic multiplicity.
On the other hand, a regular $6\times6$ quadratic matrix polynomial of grade two has twelve eigenvalues when finite and infinite eigenvalues are counted together.
Therefore, the remaining three eigenvalues are at infinity.
Accordingly, the $12$-dimensional generalized eigenvalue problem in Eq.~\eqref{eq:generalized_eigenproblem} contains
\[
    9\ {\rm finite}
    \quad+\quad
    3\ {\rm infinite}
\]
eigenvalues. 
For special parameter values, the polynomial eigenvalue problem may yield spurious eigenvalues, particularly at $z_\omega=0$. 
Such eigenvalues must be checked against the original rational matrix function $F(z_\omega)$ to determine whether they correspond to genuine poles of the Diffuson.
For example, in the $\be_z$ channel of the RD model, the cubic equation yields a spurious root at $z_\omega=0$ since $c_z=0$, see Sec.~\ref{sec:channel_decoupling_and_fixed_radius}.

\section{Application to a model with $[\mathbf{C},\mathbf{M}_1]\neq0$}
\label{app:noncommuting_model}

\begin{figure}[h]
    \centering
    \includegraphics[width=0.7\columnwidth]
    {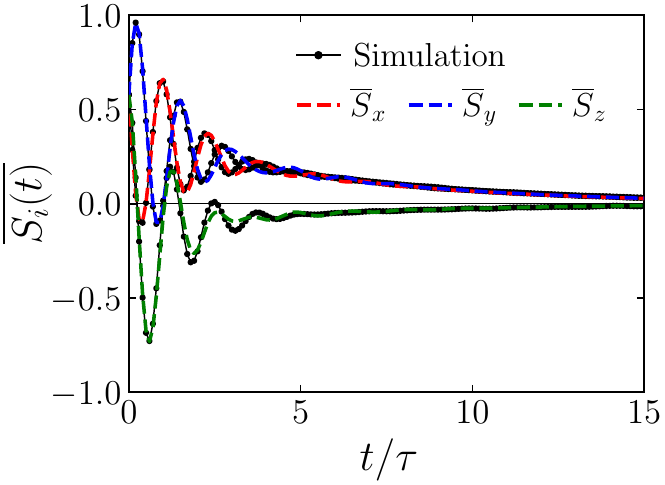}
    \caption{
    Comparison of the numerical and analytical spin dynamics for the RD model with an additional $\lambda_z k_x\sigma_z$ term.
    The coupling strength is $m\lambda_z\ell/\hbar^2=1$; all other conditions and plotting conventions are the same as for P1 in Fig.~\ref{fig:spin_dynamics}.
    }
    \label{fig:noncommuting_spin_dynamics}
\end{figure}

We consider a model for which $\mat{C}$ and $\mat{M}_1$ do not commute, so the common-eigenbasis reduction is unavailable.
Specifically, we add a term $\lambda_z k_x\sigma_z$ to the RD model of Sec.~\ref{sec:RD_model}.
We set $m\lambda_z\ell/\hbar^2=1$, with all other parameters and simulation conditions identical to those of P1 in Table~\ref{tab:params_spin_dynamics}.
As shown in Fig.~\ref{fig:noncommuting_spin_dynamics}, the analytical results obtained using the procedure of Sec.~\ref{sec:computational_procedure} agree well with the numerical simulations.

\newpage
\bibliography{refs}

@book{Akkermans_2007_book,
   author = {Akkermans, E. and Montambaux, G.},
   title = {{Mesoscopic Physics of Electrons and Photons}},
   publisher = {Cambridge University Press},
   address = {London},
   year = {2007}
}

@article{Boschini_2024_review,
  title = {{Time-resolved ARPES studies of quantum materials}},
  author = {Boschini, Fabio and Zonno, Marta and Damascelli, Andrea},
  journal = {Rev. Mod. Phys.},
  volume = {96},
  issue = {1},
  pages = {015003},
  numpages = {56},
  year = {2024},
  month = {Feb},
  publisher = {American Physical Society},
  doi = {10.1103/RevModPhys.96.015003},
  url = {https://link.aps.org/doi/10.1103/RevModPhys.96.015003}
}

@incollection{Dyakonov_2008_book,
  author    = {Dyakonov, M. I.},
  title     = {{Basics of Semiconductor and Spin Physics}},
  booktitle = {{Spin Physics in Semiconductors}},
  editor    = {Dyakonov, M. I.},
  publisher = {Springer},
  address   = {Berlin},
  year      = {2008}
}

@article{Glazov_2010_review,
   title = {{Two-dimensional electron gas with spin--orbit coupling disorder}},
   journal = {Physica E},
   volume = {42},
   number = {9},
   pages = {2157-2177},
   year = {2010},
   issn = {1386-9477},
   doi = {https://doi.org/10.1016/j.physe.2010.04.021},
   author = {M.M. Glazov and E.Ya. Sherman and V.K. Dugaev},
}

@article{Guttel_2017_review, 
   title={{The nonlinear eigenvalue problem}}, 
   volume={26}, 
   DOI={10.1017/S0962492917000034}, 
   journal={Acta Numer.}, 
   author={G\"uttel, Stefan and Tisseur, Françoise}, 
   year={2017}, 
   pages={1-94},
}

@article{Tisseur_2001_review,
   author = {Tisseur, Fran{\c c}oise and Meerbergen, Karl},
   title = {{The Quadratic Eigenvalue Problem}},
   journal = {SIAM Review},
   volume = {43},
   number = {2},
   pages = {235-286},
   year = {2001},
   doi = {10.1137/S0036144500381988},
}

@article{Wu_2010_review,
   title = {{Spin dynamics in semiconductors}},
   journal = {Phys. Rep.},
   volume = {493},
   number = {2},
   pages = {61-236},
   year = {2010},
   issn = {0370-1573},
   doi = {https://doi.org/10.1016/j.physrep.2010.04.002},
   url = {https://www.sciencedirect.com/science/article/pii/S0370157310000955},
   author = {M.W. Wu and J.H. Jiang and M.Q. Weng},
}

@article{Amundsen_2024,
  title = {{RKKY interaction in Rashba altermagnets}},
  author = {Amundsen, Morten and Brataas, Arne and Linder, Jacob},
  journal = {Phys. Rev. B},
  volume = {110},
  issue = {5},
  pages = {054427},
  numpages = {7},
  year = {2024},
  month = {Aug},
  publisher = {American Physical Society},
  doi = {10.1103/PhysRevB.110.054427},
  url = {https://link.aps.org/doi/10.1103/PhysRevB.110.054427}
}

@article{Arabahmadi_2024,
  title = {{Momentum-space signatures of the Anderson transition in a symplectic, two-dimensional, disordered ultracold gas}},
  author = {Arabahmadi, Ehsan and Schumayer, Daniel and Gr{\'e}maud, Beno{\^{\i}}t and Miniatura, Christian and Hutchinson, David A. W.},
  journal = {Phys. Rev. Res.},
  volume = {6},
  issue = {1},
  pages = {L012021},
  numpages = {6},
  year = {2024},
  month = {Jan},
  publisher = {American Physical Society},
  doi = {10.1103/PhysRevResearch.6.L012021},
  url = {https://link.aps.org/doi/10.1103/PhysRevResearch.6.L012021}
}

@article{Bernevig_2006,
  title = {{Exact SU(2) Symmetry and Persistent Spin Helix in a Spin-Orbit Coupled System}},
  author = {Bernevig, B. Andrei and Orenstein, J. and Zhang, Shou-Cheng},
  journal = {Phys. Rev. Lett.},
  volume = {97},
  issue = {23},
  pages = {236601},
  numpages = {4},
  year = {2006},
  month = {Dec},
  publisher = {American Physical Society},
  doi = {10.1103/PhysRevLett.97.236601},
  url = {https://link.aps.org/doi/10.1103/PhysRevLett.97.236601}
}

@article{Bernevig_2008,
  title = {{Semiclassical theory of diffusive-ballistic crossover and the persistent spin helix}},
  author = {Bernevig, B. Andrei and Hu, Jiangping},
  journal = {Phys. Rev. B},
  volume = {78},
  issue = {24},
  pages = {245123},
  numpages = {4},
  year = {2008},
  month = {Dec},
  publisher = {American Physical Society},
  doi = {10.1103/PhysRevB.78.245123},
  url = {https://link.aps.org/doi/10.1103/PhysRevB.78.245123}
}

@article{Burkov_2004a,
  title = {{Theory of spin-charge-coupled transport in a two-dimensional electron gas  with Rashba spin-orbit interactions}},
  author = {Burkov, A. A. and N\'u\~nez, Alvaro S. and MacDonald, A. H.},
  journal = {Phys. Rev. B},
  volume = {70},
  issue = {15},
  pages = {155308},
  numpages = {8},
  year = {2004},
  month = {Oct},
  publisher = {American Physical Society},
  doi = {10.1103/PhysRevB.70.155308},
  url = {https://link.aps.org/doi/10.1103/PhysRevB.70.155308}
}

@article{Burkov_2004b,
  title = {{Spin relaxation in a two-dimensional electron gas in a perpendicular magnetic field}},
  author = {Burkov, A. A. and Balents, Leon},
  journal = {Phys. Rev. B},
  volume = {69},
  issue = {24},
  pages = {245312},
  numpages = {6},
  year = {2004},
  month = {Jun},
  publisher = {American Physical Society},
  doi = {10.1103/PhysRevB.69.245312},
  url = {https://link.aps.org/doi/10.1103/PhysRevB.69.245312}
}

@article{Bychkov_1984,
   doi = {10.1088/0022-3719/17/33/015},
   url = {https://dx.doi.org/10.1088/0022-3719/17/33/015},
   year = {1984},
   month = {nov},
   publisher = {},
   volume = {17},
   number = {33},
   pages = {6039},
   author = {Yu A Bychkov and E I Rashba},
   title = {{Oscillatory effects and the magnetic susceptibility of carriers in inversion layers}},
   journal = {J. Phys. C},
}

@article{Cacho_2015,
  title = {{Momentum-Resolved Spin Dynamics of Bulk and Surface Excited States in the Topological Insulator ${\mathrm{Bi}}_{2}{\mathrm{Se}}_{3}$}},
  author = {Cacho, C. and Crepaldi, A. and Battiato, M. and Braun, J. and Cilento, F. and Zacchigna, M. and Richter, M. C. and Heckmann, O. and Springate, E. and Liu, Y. and Dhesi, S. S. and Berger, H. and Bugnon, Ph. and Held, K. and Grioni, M. and Ebert, H. and Hricovini, K. and Min{\'a}r, J. and Parmigiani, F.},
  journal = {Phys. Rev. Lett.},
  volume = {114},
  issue = {9},
  pages = {097401},
  numpages = {6},
  year = {2015},
  month = {Mar},
  publisher = {American Physical Society},
  doi = {10.1103/PhysRevLett.114.097401},
  url = {https://link.aps.org/doi/10.1103/PhysRevLett.114.097401}
}

@article{Clark_2022,
   author = {Clark, Oliver J. and Wadgaonkar, Indrajit and Freyse, Friedrich and Springholz, Gunther and Battiato, Marco and Sánchez-Barriga, Jaime},
   title = {{Ultrafast Thermalization Pathways of Excited Bulk and Surface States in the Ferroelectric Rashba Semiconductor GeTe}},
   journal = {Adv. Mater.},
   volume = {34},
   number = {24},
   pages = {2200323},
   doi = {https://doi.org/10.1002/adma.202200323},
   url = {https://advanced.onlinelibrary.wiley.com/doi/abs/10.1002/adma.202200323},
   year = {2022}
}

@article{Delande_2014,
  title = {{Mobility Edge for Cold Atoms in Laser Speckle Potentials}},
  author = {Delande, Dominique and Orso, Giuliano},
  journal = {Phys. Rev. Lett.},
  volume = {113},
  issue = {6},
  pages = {060601},
  numpages = {5},
  year = {2014},
  month = {Aug},
  publisher = {American Physical Society},
  doi = {10.1103/PhysRevLett.113.060601},
  url = {https://link.aps.org/doi/10.1103/PhysRevLett.113.060601}
}

@article{Dornelas_2026,
  title = {{Jittery quantum boomerang effect}},
  author = {Dornelas, Pedro and Ferreira, Gerson J.},
  journal = {Phys. Rev. B},
  volume = {114},
  issue = {14},
  pages = {L140203},
  numpages = {7},
  year = {2026},
  month = {Sep},
  publisher = {American Physical Society},
  doi = {10.1103/qxxq-y2hy},
  url = {https://link.aps.org/doi/10.1103/qxxq-y2hy}
}

@article{Dresselhaus_1955,
  title = {{Spin-Orbit Coupling Effects in Zinc Blende Structures}},
  author = {Dresselhaus, G.},
  journal = {Phys. Rev.},
  volume = {100},
  issue = {2},
  pages = {580--586},
  numpages = {0},
  year = {1955},
  month = {Oct},
  publisher = {American Physical Society},
  doi = {10.1103/PhysRev.100.580},
  url = {https://link.aps.org/doi/10.1103/PhysRev.100.580}
}

@article{Dugaev_2009,
  title = {{Spin relaxation and combined resonance in two-dimensional electron systems with spin-orbit disorder}},
  author = {Dugaev, V. K. and Sherman, E. Ya. and Ivanov, V. I. and Barna{\ifmmode \acute{s}\else \'{s}\fi{}}, J.},
  journal = {Phys. Rev. B},
  volume = {80},
  issue = {8},
  pages = {081301(R)},
  numpages = {4},
  year = {2009},
  month = {Aug},
  publisher = {American Physical Society},
  doi = {10.1103/PhysRevB.80.081301},
  url = {https://link.aps.org/doi/10.1103/PhysRevB.80.081301}
}

@article{Dyakonov_1972,
   author = {D'yakonov, M. I. and Perel', V. I.},
   title = {{Spin relaxation of conduction electrons in noncentrosymmetric semiconductors}},
   journal = {Fiz. Tverd. Tela (Leningrad)},
   volume = {13},
   pages = {3581},
   year = {1971},
   note={[Sov. Phys. Solid State \textbf{13}, 3023 (1972)]},
   type = {Journal Article}
}

@article{Eskandari-asl_2026,
  title = {{Magneto-optical Kerr effect in pump-probe setups}},
  author = {Eskandari-asl, Amir and Avella, Adolfo},
  journal = {Phys. Rev. A},
  volume = {113},
  issue = {1},
  pages = {013738},
  numpages = {13},
  year = {2026},
  month = {Jan},
  publisher = {American Physical Society},
  doi = {10.1103/wfcr-75gl},
  url = {https://link.aps.org/doi/10.1103/wfcr-75gl}
}

@article{C-Fang_2012,
  title = {{Multi-Weyl Topological Semimetals Stabilized by Point Group Symmetry}},
  author = {Fang, Chen and Gilbert, Matthew J. and Dai, Xi and Bernevig, B. Andrei},
  journal = {Phys. Rev. Lett.},
  volume = {108},
  issue = {26},
  pages = {266802},
  numpages = {5},
  year = {2012},
  month = {Jun},
  publisher = {American Physical Society},
  doi = {10.1103/PhysRevLett.108.266802},
  url = {https://link.aps.org/doi/10.1103/PhysRevLett.108.266802}
}

@Article{Glazov_2006,
   author={Glazov, M. M. and Golub, L. E.},
   title={{Nondiffusive weak localization in two-dimensional systems with spin-orbit splitting of the spectrum}},
   journal={Semiconductors},
   year={2006},
   month={Oct},
   day={01},
   volume={40},
   number={10},
   pages={1209-1217},
   issn={1090-6479},
   doi={10.1134/S1063782606100150},
   url={https://doi.org/10.1134/S1063782606100150}
}

@article{Glazov_2009,
   doi = {10.1088/0268-1242/24/6/064007},
   url = {https://doi.org/10.1088/0268-1242/24/6/064007},
   year = {2009},
   month = {may},
   publisher = {},
   volume = {24},
   number = {6},
   pages = {064007},
   author = {Glazov, M M and Golub, L E},
   title = {{Spin--orbit interaction and weak localization in heterostructures}},
   journal = {Semicond. Sci. Technol.},
}

@article{Golub_2005,
  title = {{Weak antilocalization in high-mobility two-dimensional systems}},
  author = {Golub, L. E.},
  journal = {Phys. Rev. B},
  volume = {71},
  issue = {23},
  pages = {235310},
  numpages = {6},
  year = {2005},
  month = {Jun},
  publisher = {American Physical Society},
  doi = {10.1103/PhysRevB.71.235310},
  url = {https://link.aps.org/doi/10.1103/PhysRevB.71.235310}
}

@article{Grimaldi_2005,
  title = {{Electron spin dynamics in impure quantum wells for arbitrary spin-orbit coupling}},
  author = {Grimaldi, C.},
  journal = {Phys. Rev. B},
  volume = {72},
  issue = {7},
  pages = {075307},
  numpages = {4},
  year = {2005},
  month = {Aug},
  publisher = {American Physical Society},
  doi = {10.1103/PhysRevB.72.075307},
  url = {https://link.aps.org/doi/10.1103/PhysRevB.72.075307}
}

@article{Guttel_2014,
   author = {G{\"u}ttel, Stefan and Van Beeumen, Roel and Meerbergen, Karl and Michiels, Wim},
   title = {{NLEIGS: A Class of Fully Rational Krylov Methods for Nonlinear Eigenvalue Problems}},
   journal = {SIAM J. Sci. Comput.},
   volume = {36},
   number = {6},
   pages = {A2842-A2864},
   year = {2014},
   doi = {10.1137/130935045},
   URL = {https://doi.org/10.1137/130935045},
}

@article{Hasan_2022,
  title = {{Wave Packet Dynamics in Synthetic Non-Abelian Gauge Fields}},
  author = {Hasan, Mehedi and Madasu, Chetan Sriram and Rathod, Ketan D. and Kwong, Chang Chi and Miniatura, Christian and Chevy, Fr\'ed\'eric and Wilkowski, David},
  journal = {Phys. Rev. Lett.},
  volume = {129},
  issue = {13},
  pages = {130402},
  numpages = {6},
  year = {2022},
  month = {Sep},
  publisher = {American Physical Society},
  doi = {10.1103/PhysRevLett.129.130402},
  url = {https://link.aps.org/doi/10.1103/PhysRevLett.129.130402}
}

@article{Hijano_2024,
  title = {Weak localization at arbitrary disorder strength in systems with generic spin-dependent fields},
  author = {Hijano, Alberto and Ili\ifmmode \acute{c}\else \'{c}\fi{}, Stefan and Bergeret, F. Sebasti\'an},
  journal = {Phys. Rev. Res.},
  volume = {6},
  issue = {2},
  pages = {023100},
  numpages = {12},
  year = {2024},
  month = {Apr},
  publisher = {American Physical Society},
  doi = {10.1103/PhysRevResearch.6.023100},
  url = {https://link.aps.org/doi/10.1103/PhysRevResearch.6.023100}
}

@article{Hikami_1980,
    author = {Hikami, Shinobu and Larkin, Anatoly I. and Nagaoka, Yosuke},
    title = {{Spin-Orbit Interaction and Magnetoresistance in the Two Dimensional Random System}},
    journal = {Prog. Theor. Phys.},
    volume = {63},
    number = {2},
    pages = {707-710},
    year = {1980},
    month = {02},
    issn = {0033-068X},
    doi = {10.1143/PTP.63.707},
    url = {https://doi.org/10.1143/PTP.63.707},
}

@article{Hsieh_2011,
  title = {{Selective Probing of Photoinduced Charge and Spin Dynamics in the Bulk and Surface of a Topological Insulator}},
  author = {Hsieh, D. and Mahmood, F. and McIver, J. W. and Gardner, D. R. and Lee, Y. S. and Gedik, N.},
  journal = {Phys. Rev. Lett.},
  volume = {107},
  issue = {7},
  pages = {077401},
  numpages = {5},
  year = {2011},
  month = {Aug},
  publisher = {American Physical Society},
  doi = {10.1103/PhysRevLett.107.077401},
  url = {https://link.aps.org/doi/10.1103/PhysRevLett.107.077401}
}

@Article{Huang_2016,
   author={Huang, Lianghui and Meng, Zengming and Wang, Pengjun and Peng, Peng and Zhang, Shao-Liang and Chen, Liangchao and Li, Donghao and Zhou, Qi and Zhang, Jing},
   title={{Experimental realization of two-dimensional synthetic spin--orbit coupling in ultracold Fermi gases}},
   journal={Nat. Phys.},
   year={2016},
   month={Jun},
   day={01},
   volume={12},
   number={6},
   pages={540-544},
   issn={1745-2481},
   doi={10.1038/nphys3672},
   url={https://doi.org/10.1038/nphys3672}
}

@article{Ilic_2020,
  title = {{Unified description of spin transport, weak antilocalization, and triplet superconductivity in systems with spin-orbit coupling}},
  author = {Ili{\ifmmode \acute{c}\else \'{c}\fi{}}, Stefan and Tokatly, Ilya V. and Bergeret, F. Sebasti{\'a}n},
  journal = {Phys. Rev. B},
  volume = {102},
  issue = {23},
  pages = {235430},
  numpages = {14},
  year = {2020},
  month = {Dec},
  publisher = {American Physical Society},
  doi = {10.1103/PhysRevB.102.235430},
  url = {https://link.aps.org/doi/10.1103/PhysRevB.102.235430}
}

@article{Iizasa_2018,
  title = {{Robustness of a persistent spin helix against a cubic Dresselhaus field in (001) and (110) oriented two-dimensional electron gases}},
  author = {Iizasa, D. and Sato, D. and Morita, K. and Nitta, J. and Kohda, M.},
  journal = {Phys. Rev. B},
  volume = {98},
  issue = {16},
  pages = {165112},
  numpages = {9},
  year = {2018},
  month = {Oct},
  publisher = {American Physical Society},
  doi = {10.1103/PhysRevB.98.165112},
  url = {https://link.aps.org/doi/10.1103/PhysRevB.98.165112}
}

@article{Iordanskii_1994,
   author = {Iordanskii, S. V. and Lyanda-Geller, Yu. B. and Pikus, G. E.},
   title = {{Weak localization in quantum wells with spin-orbit interaction}},
   journal = {Pis'ma Zh. Eksp. Teor. Fiz.},
   volume = {60},
   pages = {199},
   year = {1994},
   note = {[JETP Lett. \textbf{60}, 206 (1994)]},
   type = {Journal Article}
}

@article{Jendrzejewski_2012,
  title = {{Coherent Backscattering of Ultracold Atoms}},
  author = {Jendrzejewski, F. and M\"uller, K. and Richard, J. and Date, A. and Plisson, T. and Bouyer, P. and Aspect, A. and Josse, V.},
  journal = {Phys. Rev. Lett.},
  volume = {109},
  issue = {19},
  pages = {195302},
  numpages = {5},
  year = {2012},
  month = {Nov},
  publisher = {American Physical Society},
  doi = {10.1103/PhysRevLett.109.195302},
  url = {https://link.aps.org/doi/10.1103/PhysRevLett.109.195302}
}

@article{Kakoi_2024,
  title = {{Time evolution of coherent wave propagation and spin relaxation in spin-orbit-coupled systems}},
  author = {Kakoi, Masataka and Slevin, Keith},
  journal = {Phys. Rev. A},
  volume = {109},
  issue = {3},
  pages = {033303},
  numpages = {15},
  year = {2024},
  month = {Mar},
  publisher = {American Physical Society},
  doi = {10.1103/PhysRevA.109.033303},
  url = {https://link.aps.org/doi/10.1103/PhysRevA.109.033303}
}

@article{Kakoi_2026a,
  title = {{Coherent transport in two-dimensional disordered potentials under spatially uniform $\text{SU}(2)$ gauge fields}},
  author = {Kakoi, Masataka and Miniatura, Christian and Slevin, Keith},
  journal = {Phys. Rev. A},
  volume = {113},
  issue = {4},
  pages = {L041302},
  numpages = {8},
  year = {2026},
  month = {Apr},
  publisher = {American Physical Society},
  doi = {10.1103/kdcq-9q67},
  url = {https://link.aps.org/doi/10.1103/kdcq-9q67}
}

@article{Kakoi_2026b,
  title = {{Mesoscopic scattering dynamics under generic uniform SU(2) gauge fields: Spin-momentum relaxation and coherent backscattering}},
  author = {Kakoi, Masataka and Miniatura, Christian and Slevin, Keith},
  journal = {Phys. Rev. B},
  volume = {114},
  issue = {1},
  pages = {014209},
  numpages = {18},
  year = {2026},
  month = {Jul},
  publisher = {American Physical Society},
  doi = {10.1103/kyk8-71l4},
  url = {https://link.aps.org/doi/10.1103/kyk8-71l4}
}

@article{Keldysh_1951,
    author  = {Keldysh, M. V.},
    title   = {On eigenvalues and eigenfunctions of some classes of nonselfadjoint equations},
    journal = {Dokl. Akad. Nauk SSSR},
    volume  = {77},
    number  = {1},
    pages   = {11--14},
    year    = {1951},
    note    = {(in Russian)}
}

@article{Keldysh_1971,
  title={On the completeness of the eigenfunctions of some classes of non-selfadjoint linear operators},
  author={Keldysh, Mstislav Vsevolodovich},
  journal={Russ. Math. Surv.},
  volume={26},
  number={4},
  pages={15--44},
  year={1971},
  doi={10.1070/RM1971v026n04ABEH003985}
}

@article{Kikkawa_1998,
  title = {{Resonant Spin Amplification in $\mathit{n}$-Type GaAs}},
  author = {Kikkawa, J. M. and Awschalom, D. D.},
  journal = {Phys. Rev. Lett.},
  volume = {80},
  issue = {19},
  pages = {4313--4316},
  numpages = {0},
  year = {1998},
  month = {May},
  publisher = {American Physical Society},
  doi = {10.1103/PhysRevLett.80.4313},
  url = {https://link.aps.org/doi/10.1103/PhysRevLett.80.4313}
}

@Article{Kilic_2025,
   author={Kilic, Berkay and Alvarruiz, Sergio and Barts, Evgenii and van Dijk, Bertjan and Barone, Paolo and S{\l}awi{\'{n}}ska, Jagoda},
   title={{Universal symmetry-protected persistent spin textures in noncentrosymmetric crystals}},
   journal={Nat. Commun.},
   year={2025},
   month={Aug},
   day={27},
   volume={16},
   number={1},
   pages={7999},
   issn={2041-1723},
   doi={10.1038/s41467-025-63136-4},
   url={https://doi.org/10.1038/s41467-025-63136-4}
}

@Article{Koralek_2009,
   author={Koralek, J. D. and Weber, C. P. and Orenstein, J. and Bernevig, B. A. and Zhang, Shou-Cheng and Mack, S. and Awschalom, D. D.},
   title={{Emergence of the persistent spin helix in semiconductor quantum wells}},
   journal={Nature (London)},
   year={2009},
   month={Apr},
   day={01},
   volume={458},
   number={7238},
   pages={610-613},
   issn={1476-4687},
   doi={10.1038/nature07871},
   url={https://doi.org/10.1038/nature07871}
}

@article{Kohda_2012,
  title = {{Gate-controlled persistent spin helix state in (In,Ga)As quantum wells}},
  author = {Kohda, M. and Lechner, V. and Kunihashi, Y. and Dollinger, T. and Olbrich, P. and Sch\"onhuber, C. and Caspers, I. and Bel'kov, V. V. and Golub, L. E. and Weiss, D. and Richter, K. and Nitta, J. and Ganichev, S. D.},
  journal = {Phys. Rev. B},
  volume = {86},
  issue = {8},
  pages = {081306(R)},
  numpages = {5},
  year = {2012},
  month = {Aug},
  publisher = {American Physical Society},
  doi = {10.1103/PhysRevB.86.081306},
  url = {https://link.aps.org/doi/10.1103/PhysRevB.86.081306}
}

@article{Kuhn_2007,
   author = {Kuhn, R. C. and Sigwarth, O. and Miniatura, C. and Delande, D. and M\"{u}ller, C. A.},
   title = {{Coherent matter wave transport in speckle potentials}},
   doi = {10.1088/1367-2630/9/6/161},
   url = {https://doi.org/10.1088/1367-2630/9/6/161},
   year = {2007},
   month = {jun},
   publisher = {},
   volume = {9},
   number = {6},
   pages = {161},
   journal = {New J. Phys.},
}

@article{Kuroda_2016,
  title = {{Generation of Transient Photocurrents in the Topological Surface State of ${\mathrm{Sb}}_{2}{\mathrm{Te}}_{3}$ by Direct Optical Excitation with Midinfrared Pulses}},
  author = {Kuroda, K. and Reimann, J. and G{\"u}dde, J. and H{\"o}fer, U.},
  journal = {Phys. Rev. Lett.},
  volume = {116},
  issue = {7},
  pages = {076801},
  numpages = {5},
  year = {2016},
  month = {Feb},
  publisher = {American Physical Society},
  doi = {10.1103/PhysRevLett.116.076801},
  url = {https://link.aps.org/doi/10.1103/PhysRevLett.116.076801}
}

@article{Labeyrie_2012,
   doi = {10.1209/0295-5075/100/66001},
   url = {https://doi.org/10.1209/0295-5075/100/66001},
   year = {2012},
   month = {dec},
   publisher = {EDP Sciences, IOP Publishing and Società Italiana di Fisica},
   volume = {100},
   number = {6},
   pages = {66001},
   author = {Labeyrie, Guillaume and Karpiuk, Tomasz and Schaff, Jean-François and Grémaud, Benoît and Miniatura, Christian and Delande, Dominique},
   title = {{Enhanced backscattering of a dilute Bose-Einstein condensate}},
   journal = {Europhys. Lett.},
}

@Article{Leroux_2018,
   author={Leroux, F. and Pandey, K. and Rehbi, R. and Chevy, F. and Miniatura, C. and Gr{\'e}maud, B. and Wilkowski, D.},
   title={{Non-Abelian adiabatic geometric transformations in a cold strontium gas}},
   journal={Nat. Commun.},
   year={2018},
   month={Sep},
   day={04},
   volume={9},
   number={1},
   pages={3580},
   issn={2041-1723},
   doi={10.1038/s41467-018-05865-3},
   url={https://doi.org/10.1038/s41467-018-05865-3}
}

@article{ZL-Li_2024,
  title = {{Time-domain interferometry of electron weak localization through terahertz nonlinear response}},
  author = {Li, Zi-Long and Li, Xiao-Hui and Wan, Yuan},
  journal = {Phys. Rev. Res.},
  volume = {6},
  issue = {3},
  pages = {033125},
  numpages = {12},
  year = {2024},
  month = {Aug},
  publisher = {American Physical Society},
  doi = {10.1103/PhysRevResearch.6.033125},
  url = {https://link.aps.org/doi/10.1103/PhysRevResearch.6.033125}
}

@Article{Lin_2011,
   author={Lin, Y.-J. and Jim{\'e}nez-Garc{\'i}a, K. and Spielman, I. B.},
   title={{Spin--orbit-coupled Bose--Einstein condensates}},
   journal={Nature (London)},
   year={2011},
   month={Mar},
   day={01},
   volume={471},
   number={7336},
   pages={83-86},
   issn={1476-4687},
   doi={10.1038/nature09887},
   url={https://doi.org/10.1038/nature09887}
}

@article{X-Liu_2011,
  title = {{Spin dynamics in the strong spin-orbit coupling regime}},
  author = {Liu, Xin and Liu, Xiong-Jun and Sinova, Jairo},
  journal = {Phys. Rev. B},
  volume = {84},
  issue = {3},
  pages = {035318},
  numpages = {8},
  year = {2011},
  month = {Jul},
  publisher = {American Physical Society},
  doi = {10.1103/PhysRevB.84.035318},
  url = {https://link.aps.org/doi/10.1103/PhysRevB.84.035318}
}

@article{X-Liu_2012,
  title = {{Unified theory of spin dynamics in a two-dimensional electron gas with arbitrary spin-orbit coupling strength at finite temperature}},
  author = {Liu, Xin and Sinova, Jairo},
  journal = {Phys. Rev. B},
  volume = {86},
  issue = {17},
  pages = {174301},
  numpages = {14},
  year = {2012},
  month = {Nov},
  publisher = {American Physical Society},
  doi = {10.1103/PhysRevB.86.174301},
  url = {https://link.aps.org/doi/10.1103/PhysRevB.86.174301}
}

@article{Luffe_2011,
  title = {Relaxation mechanisms of the persistent spin helix},
  author = {L{\"u}ffe, Matthias C. and Kailasvuori, Janik and Nunner, Tamara S.},
  journal = {Phys. Rev. B},
  volume = {84},
  issue = {7},
  pages = {075326},
  numpages = {12},
  year = {2011},
  month = {Aug},
  publisher = {American Physical Society},
  doi = {10.1103/PhysRevB.84.075326},
  url = {https://link.aps.org/doi/10.1103/PhysRevB.84.075326}
}

@article{Mardonov_2015,
  title = {{Dynamics of Spin-Orbit Coupled Bose-Einstein Condensates in a Random Potential}},
  author = {Mardonov, Sh. and Modugno, M. and Sherman, E. Ya.},
  journal = {Phys. Rev. Lett.},
  volume = {115},
  issue = {18},
  pages = {180402},
  numpages = {5},
  year = {2015},
  month = {Oct},
  publisher = {American Physical Society},
  doi = {10.1103/PhysRevLett.115.180402},
  url = {https://link.aps.org/doi/10.1103/PhysRevLett.115.180402}
}

@article{Mukasa_2025,
   author = {Mukasa, Kohei and Masaki, Yusuke},
   title = {{Finite-momentum Superconductivity in Two-dimensional Altermagnets with a Rashba-type Spin--Orbit Coupling}},
   journal = {J. Phys. Soc. Jpn.},
   volume = {94},
   number = {6},
   pages = {064705},
   year = {2025},
   doi = {10.7566/JPSJ.94.064705},
   URL = {https://doi.org/10.7566/JPSJ.94.064705},
}

@article{Orso_2017,
  title = {Anderson Transition of Cold Atoms with Synthetic Spin-Orbit Coupling in Two-Dimensional Speckle Potentials},
  author = {Orso, Giuliano},
  journal = {Phys. Rev. Lett.},
  volume = {118},
  issue = {10},
  pages = {105301},
  numpages = {6},
  year = {2017},
  month = {Mar},
  publisher = {American Physical Society},
  doi = {10.1103/PhysRevLett.118.105301},
  url = {https://link.aps.org/doi/10.1103/PhysRevLett.118.105301}
}

@article{Oztas_2024,
   doi = {10.1088/1402-4896/ad405f},
   url = {https://doi.org/10.1088/1402-4896/ad405f},
   year = {2024},
   month = {may},
   publisher = {IOP Publishing},
   volume = {99},
   number = {6},
   pages = {065913},
   author = {Oztas, Z and Nabiollahi, O},
   title = {{Anderson localization and quenched induced dynamics of spin orbit coupled Bose--Einstein condensate in a random potential}},
   journal = {Phys. Scr.},
}

@article{Pezzoli_2013,
  title = {{Spin and energy relaxation in germanium studied by spin-polarized direct-gap photoluminescence}},
  author = {Pezzoli, Fabio and Qing, Lan and Giorgioni, Anna and Isella, Giovanni and Grilli, Emanuele and Guzzi, Mario and Dery, Hanan},
  journal = {Phys. Rev. B},
  volume = {88},
  issue = {4},
  pages = {045204},
  numpages = {12},
  year = {2013},
  month = {Jul},
  publisher = {American Physical Society},
  doi = {10.1103/PhysRevB.88.045204},
  url = {https://link.aps.org/doi/10.1103/PhysRevB.88.045204}
}

@article{Poshakinskiy_2013,
  title = {{Effect of Dresselhaus spin-orbit coupling on spin dephasing in asymmetric and macroscopically symmetric (110)-grown quantum wells}},
  author = {Poshakinskiy, A. V. and Tarasenko, S. A.},
  journal = {Phys. Rev. B},
  volume = {87},
  issue = {23},
  pages = {235301},
  numpages = {7},
  year = {2013},
  month = {Jun},
  publisher = {American Physical Society},
  doi = {10.1103/PhysRevB.87.235301},
  url = {https://link.aps.org/doi/10.1103/PhysRevB.87.235301}
}

@Article{Reimann_2018,
   author={Reimann, J. and Schlauderer, S. and Schmid, C. P. and Langer, F. and Baierl, S. and Kokh, K. A. and Tereshchenko, O. E. and Kimura, A. and Lange, C. and G{\"u}dde, J. and H{\"o}fer, U. and Huber, R.},
   title={{Subcycle observation of lightwave-driven Dirac currents in a topological surface band}},
   journal={Nature (London)},
   year={2018},
   month={Oct},
   day={01},
   volume={562},
   number={7727},
   pages={396-400},
   issn={1476-4687},
   doi={10.1038/s41586-018-0544-x},
   url={https://doi.org/10.1038/s41586-018-0544-x}
}

@Article{Reimann_2023,
   author={Reimann, J. and Sumida, K. and Kakoki, M. and Kokh, K. A. and Tereshchenko, O. E. and Kimura, A. and G{\"u}dde, J. and H{\"o}fer, U.},
   title={Ultrafast electron dynamics in a topological surface state observed in two-dimensional momentum space},
   journal={Sci. Rep.},
   year={2023},
   month={Apr},
   day={09},
   volume={13},
   number={1},
   pages={5796},
   issn={2045-2322},
   doi={10.1038/s41598-023-32811-1},
   url={https://doi.org/10.1038/s41598-023-32811-1}
}

@article{Richard_2019,
  title = {{Elastic Scattering Time of Matter Waves in Disordered Potentials}},
  author = {Richard, J{\'e}r{\'e}mie and Lim, Lih-King and Denechaud, Vincent and Volchkov, Valentin V. and Lecoutre, Baptiste and Mukhtar, Musawwadah and Jendrzejewski, Fred and Aspect, Alain and Signoles, Adrien and Sanchez-Palencia, Laurent and Josse, Vincent},
  journal = {Phys. Rev. Lett.},
  volume = {122},
  issue = {10},
  pages = {100403},
  numpages = {6},
  year = {2019},
  month = {Mar},
  publisher = {American Physical Society},
  doi = {10.1103/PhysRevLett.122.100403},
  url = {https://link.aps.org/doi/10.1103/PhysRevLett.122.100403}
}

@article{Sanchez-Barriga_2016,
  title = {Ultrafast spin-polarization control of Dirac fermions in topological insulators},
  author = {S{\'a}nchez-Barriga, J. and Golias, E. and Varykhalov, A. and Braun, J. and Yashina, L. V. and Schumann, R. and Min{\'a}r, J. and Ebert, H. and Kornilov, O. and Rader, O.},
  journal = {Phys. Rev. B},
  volume = {93},
  issue = {15},
  pages = {155426},
  numpages = {10},
  year = {2016},
  month = {Apr},
  publisher = {American Physical Society},
  doi = {10.1103/PhysRevB.93.155426},
  url = {https://link.aps.org/doi/10.1103/PhysRevB.93.155426}
}

@article{Schliemann_2003,
  title = {{Nonballistic Spin-Field-Effect Transistor}},
  author = {Schliemann, John and Egues, J. Carlos and Loss, Daniel},
  journal = {Phys. Rev. Lett.},
  volume = {90},
  issue = {14},
  pages = {146801},
  numpages = {4},
  year = {2003},
  month = {Apr},
  publisher = {American Physical Society},
  doi = {10.1103/PhysRevLett.90.146801},
  url = {https://link.aps.org/doi/10.1103/PhysRevLett.90.146801}
}

@article{Sherman_2003,
    author = {Sherman, E. Ya.},
    title = {{Random spin--orbit coupling and spin relaxation in symmetric quantum wells}},
    journal = {Appl. Phys. Lett.},
    volume = {82},
    number = {2},
    pages = {209-211},
    year = {2003},
    month = {01},
    issn = {0003-6951},
    doi = {10.1063/1.1533839},
    url = {https://doi.org/10.1063/1.1533839},
}

@article{Stanescu_2007,
  title = {Spin relaxation in a generic two-dimensional spin-orbit coupled system},
  author = {Stanescu, Tudor D. and Galitski, Victor},
  journal = {Phys. Rev. B},
  volume = {75},
  issue = {12},
  pages = {125307},
  numpages = {16},
  year = {2007},
  month = {Mar},
  publisher = {American Physical Society},
  doi = {10.1103/PhysRevB.75.125307},
  url = {https://link.aps.org/doi/10.1103/PhysRevB.75.125307}
}

@misc{Sumita_arXiv,
      title={{Density-matrix quantum kinetics of spin-mode crossover and ac Edelstein response in spin--orbit-coupled chiral metals}}, 
      author={Shuntaro Sumita and Yusuke Kato},
      eprint={2609.12936},
      archivePrefix={arXiv},
}

@Article{Szolnoki_2017,
   author={Szolnoki, L{\'e}n{\'a}rd and Kiss, Annam{\'a}ria and D{\'o}ra, Bal{\'a}zs and Simon, Ferenc},
   title={{Spin-relaxation time in materials with broken inversion symmetry and large spin-orbit coupling}},
   journal={Sci. Rep.},
   year={2017},
   month={Aug},
   day={30},
   volume={7},
   number={1SN  - 2045-2322},
   pages={9949},
   doi={10.1038/s41598-017-09759-0},
   url={https://doi.org/10.1038/s41598-017-09759-0}
}

@Article{Tao_2018,
   author={Tao, L. L. and Tsymbal, Evgeny Y.},
   title={{Persistent spin texture enforced by symmetry}},
   journal={Nat. Commun.},
   year={2018},
   month={Jul},
   day={17},
   volume={9},
   number={1},
   pages={2763},
   issn={2041-1723},
   doi={10.1038/s41467-018-05137-0},
   url={https://doi.org/10.1038/s41467-018-05137-0}
}

@article{ZY_Wang_2021,
   author = {Zong-Yao Wang  and Xiang-Can Cheng  and Bao-Zong Wang  and Jin-Yi Zhang  and Yue-Hui Lu  and Chang-Rui Yi  and Sen Niu  and Youjin Deng  and Xiong-Jun Liu  and Shuai Chen  and Jian-Wei Pan },
   title = {{Realization of an ideal Weyl semimetal band in a quantum gas with 3D spin-orbit coupling}},
   journal = {Science},
   volume = {372},
   number = {6539},
   pages = {271-276},
   year = {2021},
   doi = {10.1126/science.abc0105},
   URL = {https://www.science.org/doi/abs/10.1126/science.abc0105},
}

@article{Wenk_2010,
  title = {{Dimensional dependence of weak localization corrections and spin relaxation in quantum wires with Rashba spin-orbit coupling}},
  author = {Wenk, P. and Kettemann, S.},
  journal = {Phys. Rev. B},
  volume = {81},
  issue = {12},
  pages = {125309},
  numpages = {18},
  year = {2010},
  month = {Mar},
  publisher = {American Physical Society},
  doi = {10.1103/PhysRevB.81.125309},
  url = {https://link.aps.org/doi/10.1103/PhysRevB.81.125309}
}

@article{Weideman_1986,
   author = {Weideman, J. A. C. and Herbst, B. M.},
   title = {{Split-Step Methods for the Solution of the Nonlinear Schrödinger Equation}},
   journal = {SIAM J. Numer. Anal.},
   volume = {23},
   number = {3},
   pages = {485-507},
   year = {1986},
   doi = {10.1137/0723033},
}

@article{Z-Wu_2016,
   author = {Zhan Wu  and Long Zhang  and Wei Sun  and Xiao-Tian Xu  and Bao-Zong Wang  and Si-Cong Ji  and Youjin Deng  and Shuai Chen  and Xiong-Jun Liu  and Jian-Wei Pan },
   title = {{Realization of two-dimensional spin-orbit coupling for Bose-Einstein condensates}},
   journal = {Science},
   volume = {354},
   number = {6308},
   pages = {83-88},
   year = {2016},
   doi = {10.1126/science.aaf6689},
   URL = {https://www.science.org/doi/abs/10.1126/science.aaf6689},
}

@article{J-Xu_2021,
  title = {{Ab initio ultrafast spin dynamics in solids}},
  author = {Xu, Junqing and Habib, Adela and Sundararaman, Ravishankar and Ping, Yuan},
  journal = {Phys. Rev. B},
  volume = {104},
  issue = {18},
  pages = {184418},
  numpages = {16},
  year = {2021},
  month = {Nov},
  publisher = {American Physical Society},
  doi = {10.1103/PhysRevB.104.184418},
  url = {https://link.aps.org/doi/10.1103/PhysRevB.104.184418}
}

@misc{SM,
   note = {See Supplemental Material for details of the derivations of common eigenvectors and corresponding eigenvalues.}
}

\end{document}

\title{
Supplemental Material for
``Spin dynamics in disordered systems with generic
spin--momentum couplings: A matrix-rational formulation''
}

\author{Masataka Kakoi}
\affiliation{
Department of Physics, The University of Osaka,
Toyonaka, Osaka 560-0043, Japan
}

\maketitle

%\section*{Derivation of the coefficients listed in Table~III}
%\label{sec:SM_channel_coefficients}

In this Supplemental Material, we derive the common eigenvectors of $\mat{C}$ and $\mat{M}_1$ and their corresponding eigenvalues listed in Table~III of the main text.
We consider three models: Rashba--Dresselhaus, $d$-wave altermagnet with Rashba SOC, and 3D double-Weyl models.

\subsection{General expressions}

We define the unit field vector and the dimensionless branch splitting as
\begin{equation}
    \hat{\br}_{\bk}
    =
    \frac{\br(\bk)}{|\br(\bk)|},
    \qquad
    \Delta_{\bk}
    =
    \frac{2\tau|\br(\bk)|}{\hbar},
    \label{eq:SM_Rhat_Delta}
\end{equation}
where $\tau$ is the elastic scattering time.
The matrices $\mat{C}$ and $\mat{M}_1$ are given by
\begin{align}
    \mat{C}
    &=
    \left\langle
        \hat{\br}_{\bk}\hat{\br}_{\bk}^{\top}
    \right\rangle_{E_0},
    \label{eq:SM_C_general}
    \\
    \mat{M}_1
    &=
    \left\langle
        \frac{
            \unit_3
            -
            \hat{\br}_{\bk}\hat{\br}_{\bk}^{\top}
        }{
            1+\Delta_{\bk}^2
        }
    \right\rangle_{\!E_0},
    \label{eq:SM_M1_general}
\end{align}
where $\langle\cdots\rangle_{E_0}$ denotes the average over the constant-energy shell at $E_0$.
When $[\mat{C},\mat{M}_1]=0$, the two real symmetric matrices can be diagonalized in a common orthonormal basis $\{\be_a\}$ in pseudospin space.
The corresponding eigenvalues are
\begin{align}
    c_a
    &=
    \be_a^{\top}\mat{C}\be_a
    =
    \left\langle
        (\be_a\cdot\hat{\br}_{\bk})^2
    \right\rangle_{E_0},
    \label{eq:SM_ca_general}
    \\
    m_a
    &=
    \be_a^{\top}\mat{M}_1\be_a
    =
    \left\langle
        \frac{
            1-(\be_a\cdot\hat{\br}_{\bk})^2
        }{
            1+\Delta_{\bk}^2
        }
    \right\rangle_{\!E_0}.
    \label{eq:SM_ma_general}
\end{align}

For $|R(\bk_{\mathrm{on\text{-}shell}})|\ll\varepsilon_0(\bk_0)=\hbar^2 k_0^2/(2m)$, we can neglect the deformation of the constant-energy shell and evaluate $\br(\bk)$ at $|\bk|=k_0$ (fixed-radius shell approximation).
The mean free path is then 
\begin{equation}
    \ell
    =
    \frac{\hbar k_0}{m}\tau.
    \label{eq:SM_mean_free_path}
\end{equation}
Within this approximation, $\braket{\cdots}_{E_0}$ reduces to an angular integral.
To simplify the notation, we define
\begin{align}
    \int_{\phi}
    &:=
    \int_0^{2\pi}\frac{d\phi}{2\pi},
    \label{eq:SM_phi_integral}
    \\[1mm]
    \int_{\Omega}
    &:=
    \int_0^{2\pi}\frac{d\phi}{2\pi}
    \int_{-1}^{1}\frac{d(\cos\theta)}{2},
    \label{eq:SM_Omega_integral}
\end{align}
for a circular and a spherical shell, respectively.

\subsection{Rashba--Dresselhaus model}

We use the spin basis vectors
\begin{equation}
    \be_{\pm}
    =
    \frac{\be_x\pm\be_y}{\sqrt{2}}.
    \label{eq:SM_RD_basis}
\end{equation}
With the polar angle measured from the $(1,1)$ direction in momentum space, the field vector on the circle $k=k_0$ takes the form
\begin{equation}
    \br(k_0,\phi)
    =
    \frac{\hbar^2\kappa k_0}{m}
    \left[
        X_+(\phi)\be_+
        +
        X_-(\phi)\be_-
    \right],
    \label{eq:SM_RD_field}
\end{equation}
where
\begin{equation}
\begin{aligned}
    X_+(\phi)
    &=
    \cos\eta\sin\phi+r_3\sin3\phi,
    \\[1mm]
    X_-(\phi)
    &=
    \sin\eta\cos\phi+r_3\cos3\phi.
\end{aligned}
\label{eq:SM_RD_Xpm}
\end{equation}
Here, we assume $\alpha\geq\widetilde{\beta}\geq0$.
The off-diagonal elements of $\mat{C}$ and $\mat{M}_1$ vanish: $R_z=0$, while the remaining off-diagonal integrands are odd under $\phi\to\pi-\phi$.
The common eigenbasis is therefore $\{\be_+,\be_-,\be_z\}$.

Using the triple angle identities and setting $\mu=\sin^2\phi$, we obtain $X_\pm^2(\phi)=\chi_\pm(\mu)$, where
\begin{equation}
\begin{aligned}
    \chi_+(\mu)
    &=
    \mu
    \left[
        \cos\eta+r_3(3-4\mu)
    \right]^2,
    \\[1mm]
    \chi_-(\mu)
    &=
    (1-\mu)
    \left[
        \sin\eta+r_3(1-4\mu)
    \right]^2.
\end{aligned}
\label{eq:SM_chi}
\end{equation}
The dimensionless branch splitting is
\begin{equation}
    \Delta^2(\mu)
    =
    4(\kappa\ell)^2\chi(\mu),
    \label{eq:SM_RD_Delta}
\end{equation}
where
\begin{equation}
    \chi(\mu)=\chi_+(\mu)+\chi_-(\mu).
\end{equation}
For a function of $\mu=\sin^2\phi$, the angular integral becomes
\begin{equation}
    \int_{\phi}f(\sin^2\phi)
    =
    \frac{1}{\pi}
    \int_0^1
    \frac{d\mu}{\sqrt{\mu(1-\mu)}}
    f(\mu).
    \label{eq:SM_RD_mu_integral}
\end{equation}
Equations~\eqref{eq:SM_ca_general} and \eqref{eq:SM_ma_general} then give
\begin{align}
    c_{\pm}
    &=
    \frac{1}{\pi}
    \int_0^1
    \frac{d\mu}{\sqrt{\mu(1-\mu)}}
    \frac{\chi_{\pm}(\mu)}{\chi(\mu)},
    \label{eq:SM_RD_cpm}
    \\
    c_z
    &=0,
    \label{eq:SM_RD_cz}
\end{align}
and
\begin{align}
    m_{\pm}
    &=
    \frac{1}{\pi}
    \int_0^1
    \frac{d\mu}{\sqrt{\mu(1-\mu)}}
    \frac{
        \chi_{\mp}(\mu)
    }{
        \chi(\mu)
        [1+4(\kappa\ell)^2\chi(\mu)]
    },
    \label{eq:SM_RD_mpm}
    \displaybreak[1]
    \\
    m_z
    &=
    \frac{1}{\pi}
    \int_0^1
    \frac{d\mu}{\sqrt{\mu(1-\mu)}}
    \frac{1}{1+4(\kappa\ell)^2\chi(\mu)}.
    \label{eq:SM_RD_mz}
\end{align}
Since $R_z=0$, $1-(\be_{\pm}\cdot\hat{\br}_{\bk})^2=(\be_{\mp}\cdot\hat{\br}_{\bk})^2$, which gives the numerator $\chi_{\mp}$ in $m_{\pm}$.

\subsection{$d$-wave altermagnet with Rashba SOC}

On the circle $k=k_0$, the field vector is
\begin{equation}
    \br(k_0,\phi)
    =
    \frac{\hbar^2k_0}{m}
    \begin{pmatrix}
        \kappa_{\mathrm R}\sin\phi\\
        -\kappa_{\mathrm R}\cos\phi\\
        \kappa_{\mathrm{AM}}\cos2\phi
    \end{pmatrix}.
    \label{eq:SM_AM_field}
\end{equation}
The dimensionless branch splitting is
\begin{equation}
\begin{aligned}
    \Delta^2(\phi)
    &=4Y(\phi),
    \\[1mm]
    Y(\phi)
    &=
    (\kappa_{\mathrm R}\ell)^2
    +
    (\kappa_{\mathrm{AM}}\ell)^2\cos^22\phi.
\end{aligned}
\label{eq:SM_AM_Delta}
\end{equation}
Since $Y(\phi)$ is invariant under $\phi\to-\phi$ and $\phi\to\phi+\pi/2$, both $\mat{C}$ and $\mat{M}_1$ are diagonal in the Cartesian spin basis $\{\be_x,\be_y,\be_z\}$, with degenerate $x$ and $y$ channels.

Defining
\begin{equation}
    u
    =
    \sqrt{
        (\kappa_{\mathrm R}\ell)^2
        +
        (\kappa_{\mathrm{AM}}\ell)^2
    },
    \label{eq:SM_q}
\end{equation}
we obtain
\begin{equation}
    \int_{\phi}\frac{1}{Y(\phi)}
    =
    \frac{1}{(\kappa_{\mathrm R}\ell)u},
    \label{eq:SM_AM_u_integral}
\end{equation}
for $\kappa_{\mathrm R}>0$.
Since the angular averages of $\sin^2\phi/Y(\phi)$ and $\cos^2\phi/Y(\phi)$ are equal, the eigenvalues of $\mat{C}$ are
\begin{align}
    c_x=c_y
    &=
    \frac{\kappa_{\mathrm R}\ell}{2u},
    \label{eq:SM_AM_cxy}
    \\
    c_z
    &=
    1-\frac{\kappa_{\mathrm R}\ell}{u}.
    \label{eq:SM_AM_cz}
\end{align}
For $\mat{M}_1$, we define
\begin{align}
    J
    &:=
    \frac12 \Tr\,\mat{M}_1
    =
    \int_{\phi}\frac{1}{1+4Y(\phi)}
    \notag\\
    &=
    \frac{1}{
        \sqrt{
            [1+4(\kappa_{\mathrm R}\ell)^2]
            (1+4u^2)
        }
    },
    \label{eq:SM_AM_J}
\end{align}
and using
\begin{equation}
    \frac{1}{Y(1+4Y)}
    =
    \frac{1}{Y}-\frac{4}{1+4Y},
    \label{eq:SM_AM_partial_fraction}
\end{equation}
we obtain
\begin{align}
    m_z
    &=
    (\kappa_{\mathrm R}\ell)^2
    \int_{\phi}
    \frac{1}{Y(\phi)[1+4Y(\phi)]}
    \notag\\
    &=
    \frac{\kappa_{\mathrm R}\ell}{u}
    -
    4(\kappa_{\mathrm R}\ell)^2J.
    \label{eq:SM_AM_mz}
\end{align}
Since $\Tr\,\mat{M}_1=2J$ and $m_x=m_y$, the remaining eigenvalues are
\begin{equation}
    m_x=m_y
    =
    J-\frac{m_z}{2}.
    \label{eq:SM_AM_mxy}
\end{equation}
The resulting expressions extend continuously to $\kappa_{\mathrm R}=0$ at $\kappa_{\mathrm{AM}}\neq0$.
\\

\subsection{Three-dimensional double-Weyl model}

On the sphere $k=k_0$, the field vector is
\begin{equation}
    \br(k_0,\theta,\phi)
    =
    \frac{\hbar^2k_0}{m}
    \begin{pmatrix}
        \kappa_{\perp}\sin^2\theta\cos2\phi\\
        \kappa_{\perp}\sin^2\theta\sin2\phi\\
        \kappa_z\cos\theta
    \end{pmatrix}.
    \label{eq:SM_DW_field}
\end{equation}
With $\mu=\cos\theta$, we define
\begin{equation}
\begin{aligned}
    A(\mu)
    &=
    4(\kappa_{\perp}\ell)^2(1-\mu^2)^2,
    \\
    B(\mu)
    &=
    4(\kappa_z\ell)^2\mu^2,
    \\
    D(\mu)
    &=
    A(\mu)+B(\mu).
\end{aligned}
\label{eq:SM_DW_D}
\end{equation}
The dimensionless branch splitting is $\Delta^2(\mu)=D(\mu)$.
The azimuthal averages of $\cos^22\phi$ and $\sin^22\phi$ are $1/2$, whereas those of $\cos2\phi$, $\sin2\phi$, and $\sin2\phi\cos2\phi$ vanish.
Both $\mat{C}$ and $\mat{M}_1$ are therefore diagonal in the Cartesian spin basis, with degenerate $x$ and $y$ channels.

Performing the azimuthal integral and using the evenness of $A$, $B$, and $D$ in $\mu$, we obtain
\begin{align}
    c_x=c_y
    &=
    \int_{\Omega}
    \frac{A(\mu)\cos^22\phi}{D(\mu)}
    =
    \frac{1}{2}
    \int_0^1d\mu\,
    \frac{A(\mu)}{D(\mu)},
    \label{eq:SM_DW_cxy}
    \\
    c_z
    &=
    \int_{\Omega}
    \frac{B(\mu)}{D(\mu)}
    =
    \int_0^1d\mu\,
    \frac{B(\mu)}{D(\mu)}.
    \label{eq:SM_DW_cz}
\end{align}
Similarly,
\begin{align}
    m_z
    &=
    \int_0^1d\mu\,
    \frac{A(\mu)}{D(\mu)[1+D(\mu)]},
    \label{eq:SM_DW_mz}
    \\
    m_x=m_y
    &=
    \int_0^1d\mu\,
    \frac{1}{1+D(\mu)}
    -
    \frac{m_z}{2}.
    \label{eq:SM_DW_mxy}
\end{align}
These are the integral expressions listed in Table~III.